\documentclass[final,authoryear,5p,times,twocolumn]{elsarticle_accepted}

\usepackage{amsmath}

\usepackage{amssymb}

\usepackage{lineno}

\usepackage{rotating}

\usepackage{soul}
\usepackage[usenames,dvipsnames]{color}

\usepackage{url}

\journal{Icarus}

\newcommand{\disp}{\displaystyle}
\newcommand{\ti}{\tiny}

\newcommand{\SMM}[1]{{\color{RoyalBlue}{\textbf{#1}}}}

\newcommand{\JWB}[1]{{\color{RoyalBlue}{\textbf{#1}}}}

\newcommand{\REVfirst}[1]{{#1}} 
\newcommand{\REVsec}[1]{{#1}}  

\begin{document}
\begin{frontmatter}

\title{Structure of Titan's evaporites}

\author[GSMA,UTINAM]{D.~Cordier}
\ead{daniel.cordier@univ-reims.fr}
\author[ESAMadrid]{T.~Cornet}
\author[UIdaho]{J.~W.~Barnes}
\author[UIdaho]{S.~M.~MacKenzie}
\author[ENSLyon]{T.~Le Bahers}
\author[LPGN]{D.~Nna-Mvondo}
\author[GSMA]{P.~Rannou}
\author[UCoimbra]{A.~G.~Ferreira}
\address[GSMA]{Groupe de Spectrom\'etrie Mol\'eculaire et Atmosph\'erique - UMR 6089 
               Campus Moulin de la Housse - BP 1039
               Universit\'e de Reims Champagne-Ardenne
               51687 REIMS -- France}
\address[UTINAM]{Universit{\'e} de Franche-Comt{\'e}, Institut UTINAM, CNRS/INSU, UMR 6213, 25030 Besan\c{c}on Cedex, France}
\address[ESAMadrid]{European Space Agency (ESA), European Space Astronomy Centre (ESAC), P.O. BOX 78, 
                    E-28691 Villanueva de la Ca\~{n}ada (Madrid), SPAIN }
\address[UIdaho]{Department of Physics, University of Idaho, Engineering-Physics Building, Moscow, ID 83844, USA}
\address[ENSLyon]{Universit\'{e} de Lyon, Universit\'{e} Claude Bernard Lyon 1, ENS Lyon, Laboratoire de Chimie UMR5182, 
                  46 all\'{e}e d'Italie, 69007 Lyon Cedex 07, France}
\address[LPGN]{Laboratoire de Plan\'{e}tologie et G\'{e}odynamique LPGNantes - UMR CNRS 6112, 2, rue de la Houssini\`{e}re 
               - BP 92208 - 44322 Nantes Cedex 3 - FRANCE}                  
\address[UCoimbra]{Departamento de Engenharia Quimica, Universidade de Coimbra, Coimbra 3030-290, Portugal}
%
%

\begin{abstract} 
{Numerous geological features that could be evaporitic in origin have been identified on the 
surface of Titan. Although they seem to be water-ice poor, their main properties --chemical composition, thickness, 
stratification-- are essentially unknown. In this paper, which follows on a previous one focusing on the surface composition
\REVsec{\citep[][]{cordier_etal_2013b}}, we 
provide some answers to these questions derived from a new model.} 
{This model, based on the up-to-date thermodynamic theory known as ``PC-SAFT'', has been validated with available laboratory
measurements and specifically developed for our purpose.} 
{1-D models confirm the possibility of an acetylene and/or butane enriched central layer of evaporitic deposit. 
The estimated thickness of this acetylene-butane layer could explain the strong RADAR brightness of the evaporites. The 2-D 
computations indicate an accumulation of poorly soluble species at the deposit's margin. Among these species, 
HCN or aerosols similar to tholins could play a dominant role. Our model predicts the existence of chemically trimodal 
``bathtub rings'' which is consistent with what it is observed at the south polar lake Ontario Lacus. This work also provides plausible
explanations to the lack of evaporites in the south polar region and to the high radar reflectivity of dry lakebeds.} 
\end{abstract}

\end{frontmatter}

\section{\label{intro}Introduction}

      Among many other fascinating features, Titan, the largest satellite of Saturn, hosts lakes and seas \citep[][]{stofan_etal_2007}
likely filled by liquid hydrocarbons containing some amount of dissolved atmospheric nitrogen and various organic compounds 
\cite[][]{dubouloz_etal_1989,cordier_etal_2009,cordier_etal_2013a}.\\
      In \cite{cordier_etal_2013b} (hereafter PAP1), the authors only derived a chemical composition for the external surface of 
Titan's putative evaporites. Butane and acetylene were found to be the most likely main components of these external layers, but this 
result has several restrictions, the most obvious being the lack of information concerning the spatial structure of the evaporitic 
deposits. Indeed, in PAP1, neither vertical stratification nor horizontal variations of composition were considered; consequently 
the model can be labelled ``0-D''. The 1-D or 2-D models of evaporitic deposition are of interest as the subsequent structure is 
potentially observable at the margins of these geological units. Moreover, a future lander could drill into these layers and perform 
detailed analysis or a Titan boat could directly measure dissolved solids with a mass spectrometer \citep{stofan_etal_2011}. 

 In a laboratory study, \cite{malaska_etal_2012} obtained interesting and illustrative result on the evaporitic crystallization process
with exotic materials. After full evaporation of their working fluid (heptane at room temperature in replacement of methane and/or ethane 
in cryogenic conditions), a ``playa'' composed of the sequence of the four organic compounds initially dissolved in the liquid was left behind.
\REVsec{It should be noticed that some species used as analogs in this experimental approach do not follow the expected behavior based on their
respective solubilities. This could be explained by the specific conditions of the laboratory simulation.}
  Already in both VIMS\footnote{Visual and Infrared Mapping Spectrometer} and RADAR data, chemical composition gradients appear to surround lakebeds \citep{barnes_etal_2009a,barnes_etal_2011}.  
\cite{barnes_etal_2009a} observed several separate \textit{annuli} following the contour of the partially evaporated lake Ontario Lacus 
at Titan's south pole. \cite{moriconi_etal_2010} tentatively detected organics and nitriles in a ramp along the shore of the same lake, suggesting that
sediments and evaporites could coexist around this object.  A model of evaporite layer structure may also shed light on the possible 
cause of the relatively high RADAR reflectivity observed in dry lakebeds. Indeed, as noticed by \cite{barnes_etal_2011}, this high 
reflectivity remains unexplained and could be caused by volume scattering if the evaporite layer is at least several centimeter thick
or contains subsurface horizons \REVsec{(see also Sect.~\ref{poss1Dstruc})}.\\
      As a first step in PAP1, the Regular Solution Theory (hereafter RST) was employed to mimic the non-ideal effect in cryogenic solutions. 
Unfortunately, this approach is clearly limited \citep[][]{cordier_etal_2012}. Thus, the model of dissolution has been subtantially improved 
in this work by the use of the Perturbed-Chain Statistical Associating Fluid Theory (hereafter PC-SAFT) equation of 
state \citep[][]{gross_sadowski_2001} which is widely employed in the chemical engineering community. The PC-SAFT has 
been successfully introduced to the study of Titan by \cite{tan_etal_2013}, \cite{luspay_kuti_etal_2015} and \cite{tan_etal_2015}.
  Another improvement on the RST approach from PAP1 is the derivation of molar volumes of the relevant molecular solids from the properties 
of their crystal structure. The influence of the pressure on these volumes is moreover studied using state of the art quantum chemical calculations.
\REVsec{We emphasize that the Modified Van Laar (MLV) model developed by \cite{glein_shock_2013} belongs to the RST family 
and relies, as does our model, 
on parameters regressed on empirical data. For the only solid organic considered by \cite{glein_shock_2013}, \textit{i.e.} acetylene, we have
used the same experimental measurements, namely those published by \cite{neumann_mann_1969}.}\\
  Our paper is organized as follows. In Sec. \ref{solubilitymodel}, we describe our new PC-SAFT based model, and we give details
concerning the properties of the different molecular solids involved. We also specify the atmospheric model taken into consideration.
Sec. \ref{1Dmodel} is devoted to results obtained with our 1-D model: for a given initial state (\textit{i.e.} depth of 
liquid, assumed composition of solutes and solvents) a possible vertical structure is proposed. The question of the maximum thickness of 
evaporite deposited is also addressed. Adopting a plausible topography, in Sec. \ref{2Dmodel} we compute what could be the species 
segregation across a lakebed shore. Finally, we discuss our results and conclude in Sec. \ref{discuss} and \ref{concl}.

\section{\label{solubilitymodel}The Model of Solutes Properties}

    Although other possible sources are available in the literature, we have chosen to keep the list of studied solutes from
the work of \cite{lavvas_etal_2008a,lavvas_etal_2008b}. This has the advantage of facilitating the comparisons with previous work
(PAP1) and limits the potential sources of uncertainties which are inevitably multiplied by introducing more species. However, in the 
last section of the paper we will discuss the occurrence and the possible role of the compounds not included in our ``standard'' mixture. 
\REVsec{Although theoretical models \citep{lavvas_etal_2008a,lavvas_etal_2008b} argue in favor of their presence,
we are aware that acetylene \REVsec{has not yet been firmly detected at the surface \citep[][]{clark_etal_2010,moriconi_etal_2010}}
and that butane has not been observed in the atmosphere}.\\
Beside the \REVsec{solvents}, considered as a ternary mixture of N$_2$, CH$_4$ and C$_2$H$_6$, we therefore consider a set of six species, 
listed in Table~\ref{solids}, \REVsec{which are assumed to be deposited to the surface of Titan or extracted from the ``soil'' by cryogenic
solvents (after being previously produced in the atmosphere).}
In photochemical models 
\citep{lavvas_etal_2008a,lavvas_etal_2008b}, they reach their temperature of solidification; therefore it can 
be hypothetized that they form exotic organic snows. Once they fall to the surface of Titan, these six species 
\REVsec{(\textit{i.e.} HCN, C$_{4}$H$_{10}$, C$_{2}$H$_{2}$, CH$_{3}$CN, CO$_{2}$, C$_{6}$H$_{6}$)}
either remain in the solid state due to local conditions or \REVsec{will be} dissolved in cryogenic solvents. 
Species that have been detected by observations or produced in photochemical models but are never found at temperatures below their 
freezing point are not considered as potential lake solutes \REVsec{--this is the case for ethylene}. 
The microphysics of the formation of organic snows is ignored, although it could be the subject of interesting research in the future.
%
\begin{table}[t]
\caption[]{Solids assumed to be dissolved in the lake and some of their properties.}
\begin{center}
{\small
\begin{tabular}{lccc}
\hline
\hline 
Species         & Precipitation                        &  Melting      & Enthalpy              \\
                & rate                                 &   temperature & of melting            \\
                & {\ti molecules\,cm$^{-2}$\,s$^{-1}$} &  {\ti(K)}     & {\ti(kJ\,mol$^{-1}$)} \\
HCN             & $1.3 \times 10^{8}$$^{(a)}$          &  260.0        &  $8.406$              \\
C$_{4}$H$_{10}$ & $5.4 \times 10^{7}$$^{(a)}$          &  136.0        &  $4.661$              \\
C$_{2}$H$_{2}$  & $5.1 \times 10^{7}$$^{(a)}$          &  192.4        &  $4.105$              \\
CH$_{3}$CN      & $4.4 \times 10^{6}$$^{(a)}$          &  229.3        &  $6.887$              \\
CO$_{2}$        & $1.3 \times 10^{6}$$^{(a)}$          &  216.6        &  $9.020$              \\
C$_{6}$H$_{6}$  & $1.0 \times 10^{6}$$^{(b)}$          &  279.1        &  $9.300$              \\
\hline
\end{tabular}
}
\\{\small$^{(a)}$\cite{lavvas_etal_2008a,lavvas_etal_2008b}; $^{(b)}$\cite{vuitton_etal_2008}.}
\end{center}
\label{solids}
\end{table}

   In the next section, we describe the adopted solubility theory and the method employed to get reliable molar volumes for organic solids.

\subsection{\label{modsolub}The Model of Solubility}

   Similar to what has been done in PAP1, our solubility estimations are made by solving the equation
\begin{equation}\label{EQbase}
 \mathrm{ln}\, \Gamma_{i} \, X_{i, sat} = -\frac{\disp\Delta H_{i,m}}{\disp R T_{i,m}} \, 
                     \left(\frac{T_{i,m}}{T}-1\right)
\end{equation}
where $X_{i, sat}$ is the mole fraction of the compound $i$ at saturation and $\Gamma_{i}$ is the activity coefficient of the considered
species. $T_{i,m}$ and $\Delta H_{i,m}$ are  melting temperature and  enthalpy of melting respectively. The temperature of the system 
is denoted $T$, and $R$ is the gas constant. This relation can be found, for instance, in the textbook by \cite{poling_2007}. The physical 
meaning of Eq. (\ref{EQbase}) is that a thermodynamic equilibrium between the considered precipitated solid $i$ and the liquid solution 
-- Eq. (\ref{EQbase}) is nothing more than an equality of chemical potential. We emphasize that $X_{i} < X_{i, sat}$ can easily 
occur for a stable state, while situations where $X_{i} > X_{i, sat}$ are metastable. Commonly, metastable states are not 
sustainable: any perturbation ignites crystallization and the corresponding mole fractions are adjusted such as $X_{i}= X_{i, sat}$. 
The overabundance of species $i$ is deposited at the bottom of the system. In PAP1 \REVsec{and in \cite{glein_shock_2013}}, 
the limitation of the validity of Eq.~(\ref{EQbase}) 
is mentioned, an in depth discussion of that aspect will be put forward in the \REVsec{appendix} of this paper.
  \cite{cordier_etal_2012} have shown the flaws of the RST, as have other authors \citep[][]{glein_shock_2013}. At its core, the RST 
is a generalization of a model established for binary mixtures. The main caveat concerning the RST probably lies in its weak physical 
foundation. In contrast, the equation of state (EoS) called PC-SAFT\footnote{Perturbed-Chain Statistical Associating Fluid Theory} 
\cite[][]{gross_sadowski_2001}, which belongs to the vast family of the SAFT EoS, is molecular based. 
\REVfirst{Indeed, PC-SAFT is derived, contrary to the RST, from the statistical physics. Each type of molecule is represented by parameters 
related to its individual microscopic properties. In that sense, PC-SAFT can be considered more profound than theories belonging to the RST
family.}
Furthermore, \REVfirst{PC-SAFT} has proved to be 
one of the most powerful types of EoS for the liquid and vapor states. \REVfirst{This theory} is the subject of numerous works in the field of thermophysics.
Here, the activity coefficient $\Gamma_{i}$ that appears in Eq. (\ref{EQbase}) will be computed with the help of PC-SAFT. 
For this application to solid-liquid equilibrium (SLE), the activity coefficient is written as the ratio $\Gamma_{i}= \Phi_{i}^{\rm L}/\Phi_{i}^{\rm L0}$, 
where $\Phi_{i}^{\rm L}$ is the fugacity coefficient of the species $i$ and $\Phi_{i}^{\rm L0}$ is the fugacity coefficient of the pure 
subcooled liquid of the same compound. In the frame of PC-SAFT, molecules are considered as ``chains'' of segments where each molecule 
is characterized by its pure-component parameters: the number of segments $m$, the segment diameter $\sigma$ (\AA) and the segment 
energy of interaction $\epsilon/k_{\rm B}$ (K). The PC-SAFT is extended to mixtures using the Berthelot-Lorentz combining rule for 
the dispersive energy, resulting in a single binary parameter $k_{ij}$. The values of all these parameters are determined by comparison 
with experimental results.
%
%
%
\begin{figure}[!t]
\includegraphics[angle=-90, width=9.5cm]{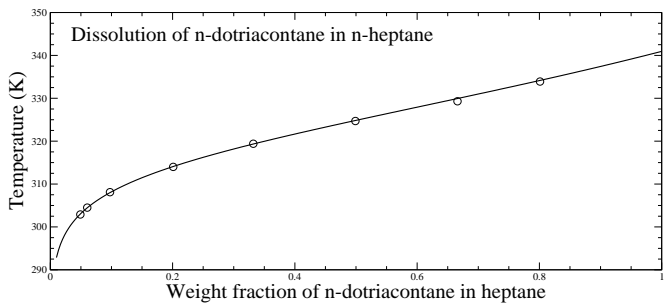}
\caption[]{\label{dotriacontane_heptane}Open circles: experimental solubility data for n-dotriacontane in n-heptane from
          \cite{chang_etal_1983}. Solid line: our model.}
\end{figure}
%
  Our implementation of PC-SAFT consists of a set of \verb+FORTRAN 2008+ object-oriented subroutines written from scratch. Our model 
has been validated in two ways. We compared its outputs with experimental unitary and binary mixtures data for vapor-liquid equilibrium (hereafter VLE), 
largely similar to those already used by \cite{tan_etal_2013} with the exception of CH$_4$-C$_2$H$_6$ mixtures \citep[see][]{luspay_kuti_etal_2015}. 
We also checked that SLE results were in good agreement with laboratory measurements (similar to the work of \cite{maity_2003}).
%
%
%
\begin{figure}[!t]
\includegraphics[angle=-90, width=9.5 cm]{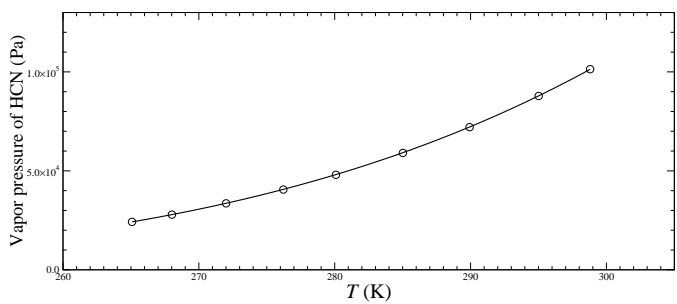}
\caption[]{\label{paramHCN}Open circles: vapor pressure for a VLE of HCN, these experimental data are from Table II of 
           \cite{giauque_ruehrwein_1939}. Solid line: our best fit leading to the PC-SAFT parameters:
           $m= 2.434$, $\sigma= 3.2929$ \AA{ } and $\epsilon/k_{\rm B}= 248.48$ K.}
\end{figure}
%
For instance we verified that we correctly reproduced the data of dissolution of dotriacontane 
\REVsec{in heptane at varying temperatures (see Fig.~\ref{dotriacontane_heptane}).}
  Our pure-component parameters are mainly taken from \cite{tan_etal_2013}, who used their own optimized adjustments. The NIST database 
or other sources, like the PC-SAFT original paper by \cite{gross_sadowski_2001}, complete the sample. \REVfirst{Concerning CH$_{3}$CN, we used 
the parameters published by \cite{spuhl_etal_2004} and decided to neglect the association terms in the Helmholtz energy since they 
only provide improvements of the order of a few percent \citep[see for instance Table 7 of][]{spuhl_etal_2004}.
This correction remains small compared to the other uncertainties related to the present modelling approach (exact composition of the solvent, influence
of the interaction parameter and validity\footnote{see \ref{append}} of Eq.~\ref{EQbase}). Additionally, 
solubilities are very sensitive
to the value of the interaction parameters $k_{ij}$ which are not known for nitriles relevant to this study \citep[][]{stevenson_etal_2015b}.} 
We did not find \REVfirst{$m$, $\sigma$ and $\epsilon/k_{\rm B}$} for HCN in the literature. 
Thus, we determined our own values by fitting the VLE data published by \cite{giauque_ruehrwein_1939} (see their Table II).
Our adjustement is compared to data from \cite{giauque_ruehrwein_1939} in Fig~\ref{paramHCN}.
%
\begin{table}[htbp]
\caption[]{The PC-SAFT pure-component parameters used in this study.}
\begin{center}
{\tiny
\begin{tabular}{lllll}
\hline
Name          & $m$       & $\sigma$ (\AA) & $\epsilon/k_{\rm B}$ (K) & References \\
CH$_4$        & 1.000     & 3.7039         & 150.030                  & {\tiny NIST, used by \cite{tan_etal_2013}}\\
N$_2$         & 1.2414    & 3.2992         & 89.2230                  & {\tiny NIST, used by \cite{tan_etal_2013}}\\
C$_2$H$_6$    & 1.6114    & 3.5245         & 190.9926                 & {\tiny NIST, used by \cite{tan_etal_2013}}\\
HCN           & 2.434     & 3.2929         & 248.48                   & {\tiny This work}\\
C$_4$H$_{10}$ & 2.6300    & 3.5100         & 190.900                  & {\tiny\cite{tamouza_2004}}\\
C$_2$H$_2$    & 2.1569    & 2.9064         & 168.5506                 & {\tiny\cite{din_1962} used by \cite{tan_etal_2013}}\\
CH$_3$CN      & 2.2661    & 3.3587         & 313.04                   & {\tiny\cite{spuhl_etal_2004}}\\
CO$_2$        & 2.0729    & 2.7852         & 169.210                  & {\tiny\cite{gross_sadowski_2001}}\\
C$_6$H$_6$    & 2.4653    & 3.6478         & 287.350                  & {\tiny\cite{gross_sadowski_2001}}\\ 
\hline
\end{tabular}
}
\end{center}
\label{MSE}
\end{table}
%
   All the $m$, $\sigma$, and $\epsilon/k_{\rm B}$ values used in this study are summarized in Table~\ref{MSE}. PC-SAFT also needs
interaction parameters $k_{ij}$ to account for interspecies molecular interaction, which may not be included in the adopted expression 
of the Helmholtz energy. In general, these $k_{ij}$ are derived from VLE experimental data and are related to binary mixtures. 
Table~\ref{PI} summarizes the interaction parameters adopted here. It should be noted that C$_2$H$_2$, CO$_2$\SMM{,} and C$_6$H$_6$ 
parameters have been derived from laboratory measurements published by \cite{neumann_mann_1969}, \cite{cheung_zander_1968} and 
%
%
%
\begin{figure}[!t]
\begin{center}
\vspace{-1cm}
\includegraphics[angle=0, width=10 cm]{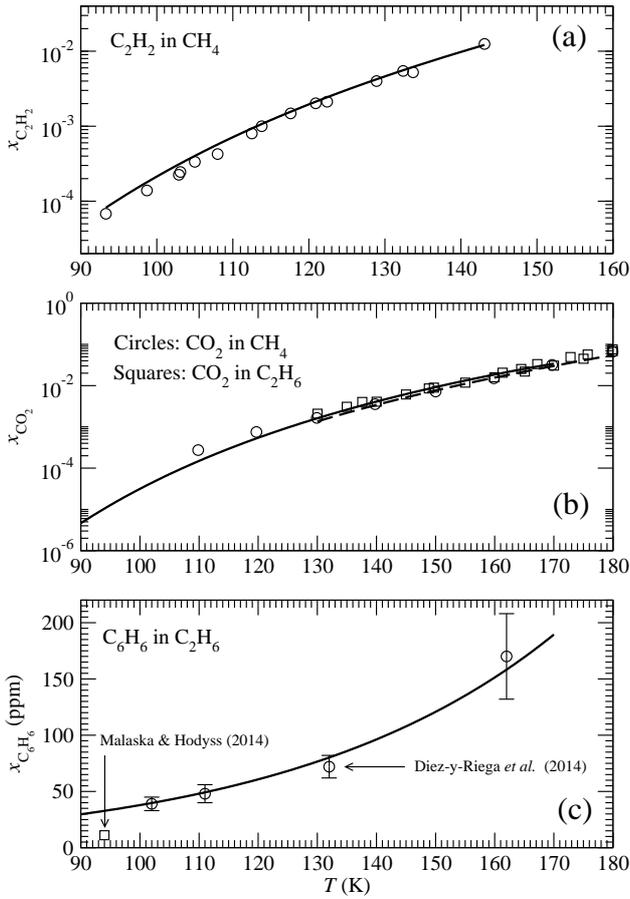}
\caption[]{\label{compaPCSAFTexpe}\small Comparison between our solubility model and experimental data: the solubility of the considered species 
           in a solvent (CH$_4$ or C$_2$H$_6$) is plotted as a function of temperature. (a) Mole fraction of dissolved acetylene in methane, 
           the observed values have been taken in \cite{neumann_mann_1969} and the solid line represents our model. (b) The measured mole 
           fraction of CO$_2$ dissolved in methane (circles) and in ethane (squares). (c) Dissolution of benzene (in ppm) in ethane, 
           experimental data are from \cite{diezyriega_etal_2014} (circles) and
           \cite{malaska_hodyss_2014} (the square at 94 K).}
\end{center}
\end{figure}
%
\cite{diezyriega_etal_2014}, respectively. The rather good agreement between our own model and experimental data is shown in 
Fig.~\ref{compaPCSAFTexpe}. 
%
%
\REVsec{Concerning the dissolution of C$_6$H$_6$ in ethane, we recognize that the measured value by \cite{malaska_hodyss_2014} at $94$ K
disagrees somewhat with those determined by \cite{diezyriega_etal_2014} (see Fig.~\ref{compaPCSAFTexpe}. c) 
but we used the measurements derived from \cite{diezyriega_etal_2014} since they were acquired over a temperature range and provide the 
necessary parameters for our PC-SAFT model.}
%
%
All unavailable interaction parameters have been set to zero.
%
\begin{table}[htbp]
\caption[]{The PC-SAFT binary interaction parameters $k_{ij}$. Only interactions between solute molecules and solvent ones have been 
taken into consideration in our model. By default, in the cases where dissolution data are not present in the literature, $k_{ij}$ 
have been fixed to zero. For C$_6$H$_6$-C$_2$H$_6$, the best fit has been derived for a temperature dependent parameter. 
Solute-to-solute molecule interactions are ignored; this assumption can be considered a relatively safe assumption since solute
abundances remains relatively low.}
\begin{center}
{\small
\begin{tabular}{llll}
\hline
Name          & CH$_4$      & N$_2$       & C$_2$H$_6$                              \\
CH$_4$        & 0           & 0.03 (1)    & 0.00 (1)                                \\
N$_2$         &             & 0           & 0.06 (2)                                \\
C$_2$H$_6$    &             &             & 0                                       \\
\hline
\hline
HCN           & 0 (X)       & 0 (X)       & 0 (X)                                   \\
C$_4$H$_{10}$ & 0.022 (4)   & 0 (X)       & 0 (X)                                   \\
C$_2$H$_2$    & 0.115 (3)   & 0 (X)       & 0.105 (1)                               \\
CH$_3$CN      & 0 (X)       & 0 (X)       & 0 (X)                                   \\
CO$_2$        & 0.085 (5)   & 0 (X)       & 0.13 (5)                                \\
C$_6$H$_6$    & 0.037 (4)   & 0 (X)       & $-0.1388 + 15.070 \times 10^{-4} \, T$ (6) \\ 
\hline
\end{tabular}
}
\end{center}
{\small(1) \cite{tan_etal_2013}; (2) this work, by fitting data from \cite{gabis_1991} provided by \cite{glein_shock_2013};
(3) this work, by fitting data from \cite{neumann_mann_1969}; (4) \cite{gross_sadowski_2001}; (5) this work, by fitting \cite{cheung_zander_1968} 
and previously used by \cite{preston_prausnitz_1970}; (6) this work, by fitting data from \cite{diezyriega_etal_2014};(X) set to zero, 
as dissolution data were not found in the literature.}
\label{PI}
\end{table}
%
%
\begin{table}[!t]
\caption[]{Cell parameters of the crystal of the different molecular solids, observed to be stable at 
           the temperature range of Titan's surface, 90K-95K. The temperature listed in the fifth column is used to experimentally
           determine the cell parameters.}
\begin{center}
{\tiny
\begin{tabular}{llllll}
\hline
Name          & a / \AA	& b / \AA	& c / \AA	& $\beta$ / $^{\rm o}$	& Ref \\
HCN           & 4.13      & 4.85      & 4.34    & 90	                & {\tiny\citep[][]{dietrich_etal_1975,dulmage_lipscomb_1951}}\\
C$_4$H$_{10}$ & 4.1463    & 7.629     & 8.169   & 118.656               & {\tiny\citep[][]{refson_pawley_1986}}\\
C$_2$H$_2$    & 6.198     & 6.023     & 5.578	& 90                    & {\tiny\citep[][]{mcmullan_etal_1992}}\\
CH$_3$CN      & 6.05      & 5.24      & 7.79    & 90                    & {\tiny\citep[][]{antson_etal_1987}}\\
CO$_2$        & 5.624     & 5.624     & 5.624   & 90                    & {\tiny\citep[][]{etters_kuchta_1989,simon_peters_1980}}\\
C$_6$H$_6$    & 7.384     & 9.416     & 6.757   & 90                    & {\tiny\citep[][]{craven_etal_1993}}\\ 
\hline
\end{tabular}
}
\end{center}
\label{table_1_Tangui}
\end{table}
%
\subsection{\label{molvolorg}The Model for the Evaporite Layers Thickness}

  The above-described model of liquid solutions provides only the number of moles of the various involved compounds that \REVsec{precipitate}.
Here we aim to estimate the thickness of the deposited layers; for this purpose, we thus need a model for the molar volumes of relevant species.\\
  If during the time step $\Delta t$ the precipitated quantities of organic matter are $\Delta n_{i}$ \REVsec{(in mol)}, the resulting thickness $\Delta e$ 
\REVsec{(in m)}
of the layer deposited over one square meter is given by $\Delta e= \sum_{i=1}^{N_{sat}} \Delta n_{i} \times V_{i,m}$. In this equation 
$N_{sat}$ denotes the number of species that reach saturation during $\Delta t$, and $V_{i,m}$  represents the molar volume 
\REVsec{(in m$^{3}$.mol$^{-1}$)} of the solid $i$. 
\REVfirst{It should be noted that inter-species possible interactions that could induce a deviation from the molar volume additivity
and/or the problem of mechanical compaction are neglected.} Hence, organic matter is 
assumed to form monocrystal structures, leaving no empty spaces in the evaporite layer as could be found in a porous medium. In that sense, 
the thicknesses calculated here are minimum values.\\
%
\begin{table}[t]
\caption[]{The densities and derived molar volumes of the crystal structures of the different molecular solids, observed stable in 
           the temperature range 90K-95K.}
\begin{center}
{\small
\begin{tabular}{llll}
\hline
              &  $\rho_{\rm solid}$  & Molar mass     & $V_{i,m}$               \\
Name          &  (g.cm$^{-3}$)       & (g.mol$^{-1}$) & (m$^{3}$.mol$^{-1}$)    \\
HCN           &  1.03                & 27.0253        & $2.624 \times 10^{-5}$  \\
C$_4$H$_{10}$ &  0.851               & 58.1222        & $6.830 \times 10^{-5}$  \\
C$_2$H$_2$    &  0.831               & 26.0373        & $3.133 \times 10^{-5}$  \\
CH$_3$CN      &  1.10                & 41.0519        & $3.732 \times 10^{-5}$  \\
CO$_2$        &  1.643               & 44.0095        & $2.679 \times 10^{-5}$  \\
C$_6$H$_6$    &  1.104               & 78.1118        & $7.075 \times 10^{-5}$  \\ 
\hline
\end{tabular}
}
\end{center}
\label{table_molvol}
\end{table}
  The molar volumes employed here are derived from the lattice parameters of the crystal cells of the organic compounds. 
For species where different crystal structures were experimentally observed, we chose the one stable at the temperature conditions 
found at the surface of Titan (\textit{i.e.} 90 K - 95 K). 
Table~\ref{table_1_Tangui} brings together the crystal structures used in our model,
which were measured at Earth ground atmospheric pressure. 
\REVsec{In the case where}
the volumes of the crystal cells were published for different temperatures, we verified that the influence of temperature 
variations on derived molar volumes is small enough to be ignored in our range of interest. The possible influence of pressure
should be weak as the pressure at Titan's surface is $\sim 1.5$ bar; we evaluated its influence by means of a  
quantum chemical calculations along with the density functional theory (DFT).
We found in general that the density of organic crystals
decreases less than 1\% over pressure ranges from $1$ bar to $100$ bar. C$_2$H$_2$, C$_6$H$_6$ and C$_4$H$_{10}$ are exceptions
for which the respective decreases were of $-1.21$\%, $-4.76$\% and $-8.46$\%. We therefore conclude, as expected, that pressure 
variations are minor factors in the context of our work. The adopted molar volumes are listed in Table~\ref{table_molvol},
\REVsec{slight differences between these molar volumes and those published in \cite{cornet_etal_2015} are explained by a tentative correction 
to account for the temperature influence on the molar volumes in their paper (given at 91.5 K instead of 90 K). In addition, the difference in 
the mass density of CO$_2$ between the two studies is due to an error in the molar mass used for conversion to density
(reported as 40 g.mol$^{-1}$ instead of 44 g.mol$^{-1}$).}
%
\section{\label{1Dmodel}Evaporite Structure: a 1D Model}
%
\subsection{Evaporite Formation Scenario}
%
%
%
%
\begin{figure}[!t]
\begin{center}
\hspace{-0.5cm}\includegraphics[angle=-90, width=9.5 cm]{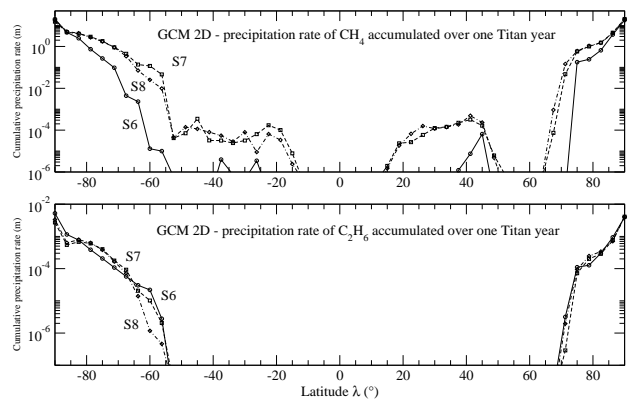}
\caption[]{\label{tauxprecipRannou}(a) The cumulatative precipitations (in meters) of methane as a function of Titan's latitude, 
           computed by \cite{rannou_etal_2006}. (b) The same quantity for ethane. In both cases, labels S6, S7, and S8 correspond 
           to various assumptions concerning the GCM 2D model inputs (\textit{e.g.} thermal inertia and lake fraction at poles).}
\end{center}
\end{figure}
%
%
   The formation of any evaporite layer requires a sequence of wet and dry periods. During the wet episode, methane and/or ethane 
rains dissolve the solid organics encountered along their runoff at the ground, and then finally they fill the 
lacustrine depressions. The subsequent dry period produces the evaporation of the solvents, and thus the formation 
of evaporites.  
The resulting vertical distribution of species depends on both the initial composition of the mixture and the individual concentrations at saturation. 
In this context, the precipitation of solid organics from the atmosphere in the form of exotic snows or hails 
has been supposed to have either taken place prior to the flowing episode or have happened at the same time as the \REVsec{runoff}. 
However, any solid organic atmospheric precipitation that occurs during the evaporation process would complicate the global picture 
of the evaporitic layer formation as it would increase the abundances of certain dissolved species. 
      The production of organics at the surface, or even in the satellite interior, cannot be excluded, although clear evidence
for such processes are not available. These still speculative phenomena could provide an organic stratum prior to any rainfall.\\

   The sequence of dry-wet periods can span over just a single year if driven by Titan's seasonal effect. Alternatively, the formation 
of observed putative evaporites observed by \cite{barnes_etal_2011} and \cite{macKenzie_etal_2014} could be the consequence of the climate 
change over much longer timescales. However, the map of evaporite distribution published by \cite{macKenzie_etal_2014} (see their Fig. 2).
is globally consistent with the latitudinal distribution of methane rains obtained by \cite{rannou_etal_2006} (see our Fig.~\ref{tauxprecipRannou}).
The largest number of deposits is concentrated in polar regions where the 
\REVsec{highest cumulative rainfall is predicted to occur}, whereas the 5-$\mu$m-bright material
detected by \cite{macKenzie_etal_2014} in the equatorial region (\textit{i.e.} around $\lambda \sim -$3$0^{\rm o}$) is consistent 
with the low, but non-zero, methane precipitation rates found by \cite{rannou_etal_2006} \REVsec{and observational evidence
\cite[][]{turtle_eal_2011a,barnes_etal_2013}}. 
%
%

%
  Following \REVsec{Rannou \textit{et al.}'s} numerical simulations, the precipitation rate of CH$_4$ poleward from $\pm 80^{\rm o}$, ranges between 
$1$ and $20$ m per Titan year. In the case of ethane, this rate is much lower with values ranging from $3 \times 10^{-4}$ 
to $5 \times 10^{-3}$ m per Titan year. These numbers yield formation times for a column of liquid methane with a height of $100$ m\
to be between $5$ and $100$ Titan years. Much longer periods of time are needed for ethane: the accumulation of a column of $100$ m 
of ethane would take \REVsec{$20,000$ - $300,000$ Titan years}.\\
  However, these timescales of lake replenishment have to be considered as an upper limit because for one particular lake, as on
Earth, the liquid catchment area (\textit{i.e.} the drainage basin) is much more extended than the lakebed (\textit{i.e.} the
lacustrine depression) itself. For instance, an examination of Fig. 1 of \cite{barnes_etal_2011} allows 
the reader to see that a collecting area with a surface ten (or even more) times that of the bed is quite common. The larger the 
surface of the drainage basin is, the shorter the lakebed replenishment time is. With a relatively large drainage basin, a replenishment
timescale shorter than one Titan year is plausible.\\
      In this work, we adopt a ``standard" initial liquid depth of $100$ m \REVsec{which}
 is broadly consistent with the bathymetry of Ligeia Mare derived
from altimetry measurements by \cite{mastrogiuseppe_etal_2014}. Besides, the computed evaporite layer structures could be easily rescaled 
to other initial liquid depths by applying a simple rule of proportionality. 
For example, a lake initially filled only with $10$ m would correspond to a final layer tenfold thinner.\\
%
%
%
%
    In the last decade, the application of GCMs has contributed much to our understanding of Titan's climate history and evolution.
However, it is not straightforward, if even possible, to quantitatively compare all published GCM results because assumptions associated 
with each model differ significantly from one study to another. For instance, the microphysics required to approach realistic precipitation 
rates have been implemented in only two investigations \citep[][]{tokano_etal_2001,rannou_etal_2006}; the radiative transfer could be 
based either on a two-stream model \citep[e.g. GCM by][]{rannou_etal_2006,tokano_2009b} or on a gray atmosphere 
\citep[e.g. GCM by][]{mitchell_etal_2011,schneider_etal_2012}; the dimensionality is either $2$ or $3$; and the methane reservoir could be 
considered finite or infinite. Most studies focus mainly on the predicted cloud activity, which presents the advantage of applicability to
observational constraints. Unfortunately, the reported simulations cannot be taken at face value to derive cumulative methane/ethane 
precipitation rates. If we look at one of the most recent works, \textit{i.e.} \cite{schneider_etal_2012}, and more specifically 
to their Fig. 1b in which the net evaporation rate $(E-P)$ has been plotted, a succession of wet and dry periods can be clearly seen
at Titan's poles. Nevertheless, polar dry periods appear to undergo evaporation rates $(E-P) \sim 0$ mm.day$^{-1}$, whereas slightly 
positive $(E-P)$'s seem to occur at latitudes around $\pm 30^{\rm o}$. At the first glance, this seems to be in contradiction with 
the evaporite distribution found by \cite{macKenzie_etal_2014}. However, \cite{schneider_etal_2012} hypothesized a total methane 
content equivalent to $12$ m of global liquid methane, and their results are averaged over $25$ Titan years, which could erase the 
temporal fluctuations.\\
  If we accept all the reservations mentioned above, and if we take cumulative methane precipitation rates computed by \cite{rannou_etal_2006}, 
then the existence (at least ephemerally) of local topological depressions filled by several tens of meters deep liquid methane looks plausible. 
Though the model is substantially different (microphysics is not included and the methane reservoir is finite), results from \cite{schneider_etal_2012} 
lead to a similar conclusion.
%
%
  \cite{mitri_etal_2007} have proposed estimations of hydrocarbon evaporation rates based on an equation originally published by 
\cite{fairall_elat_1996}. For instance, \REVsec{they} found for a pure methane liquid layer an evaporation rate of $\sim 5 \times 10^{3}$ kg.m$^{-2}$.yr$^{-1}$; 
a value that yields to $\sim 10$ terrestrial years for the complete evaporation of a column of $100$ m of liquid methane. 
This estimation is clearly compatible with the duration of a Titan season (\textit{i.e.} $\sim 7$ Earth years) and allows the evaporation 
of a transient methane lake within that time period. The same authors obtained an evaporation rate of $1.5 \times 10^{3}$ kg.m$^{-2}$.yr$^{-1}$
for a mixture of $35$\% of CH$_4$, $60$\% of ethane and $5$\% of nitrogen, an acceptable range. We note that \cite{tokano_2009} used 
the same prescription for his limnological study. Under slightly different condition\SMM{s} (no wind), \cite{luspay_kuti_etal_2012b,luspay_kuti_etal_2015} 
obtained similar evaporation rates between $\sim 0.5 \times 10^{-4}$ and 
$\sim 2 \times 10^{-4}$ kg.m$^{-2}$.s$^{-1}$ (equivalent to $1.58$--$6.31 \times 10^{3}$ kg.m$^{-2}$.yr$^{-1}$) in experimental simulations,
depending on the actual content of ethane. In any case, the evaporation of several tens of meters deep hydrocarbon lake is likely 
possible within a few Titan months. Hence, the formation of at least a thin layer of evaporite is compatible with what we know about 
evaporation rates\REVsec{, and it is plausible that the formation of the evaporites observed by Cassini have occured during 
the Titan's past.} Some authors consider alternative mechanisms for the formation of features that are classified as ``evaporites",
which we discuss in more detail in Sec. \ref{discuss}.
%
%
%
\begin{table*}[!t]
\caption[]{The solubility (in mole fraction) of considered solutes at $T= 90$ K and under $1.5$ bar for an ideal solution and with our 
           PC-SAFT based model. The solvent is only composed of either methane or ethane, $e_{100}$ represents the final thickness of
           evaporites after the evaporation of an initial column of $100$ m of liquid. These thicknesses were computed with PC-SAFT
           model solubilities. The notation $x \times 10^{y} = x \, (y)$ is used for conveniency.}
\begin{center}
{\small
\begin{tabular}{llllll}
\hline
              &  Ideal                 & PC-SAFT               & PC-SAFT               & $e_{100}$ (m)         & $e_{100}$ (m)  \\
Name          &  solution              & pure CH$_4$           & pure C$_2$H$_6$       & pure CH$_4$           & pure C$_2$H$_6$ \\
HCN           &  $6.46 \, (-4)$ & $3.52 \, (-7)$ & $4.64 \, (-5)$ & $2.59 \, (-5)$ & $2.65 \, (-3)$\\
C$_4$H$_{10}$ &  $1.26 \, (-1)$ & $1.67 \, (-3)$ & $9.14 \, (-2)$ & $3.20 \, (-1)$ & $13.6$\\
C$_2$H$_2$    &  $5.40 \, (-2)$ & $4.84 \, (-5)$ & $5.21 \, (-4)$ & $4.26 \, (-3)$ & $3.56 \, (-2)$\\
CH$_3$CN      &  $3.73 \, (-3)$ & $4.27 \, (-8)$ & $1.87 \, (-5)$ & $4.47 \, (-6)$ & $1.55 \, (-3)$\\
CO$_2$        &  $8.72 \, (-4)$ & $2.45 \, (-6)$ & $4.37 \, (-6)$ & $1.84 \, (-4)$ & $2.55 \, (-4)$\\
C$_6$H$_6$    &  $2.20 \, (-4)$ & $7.20 \, (-9)$ & $2.97 \, (-5)$ & $1.43 \, (-6)$ & $4.58 \, (-3)$\\
\hline
Total  thickness     &(m)              &                       &                       & 0.324 & 13.65\\
\end{tabular}
}
\end{center}
\label{saturationSOLUB}
\end{table*}
%
%

%
\begin{table}[!t]
\caption[]{The initial mixtures of solutes taken into account. The entire set of compositions is divided into four types. 
 Type A and B are used with a methane rich solvent whereas type C and and D correspond to when ethane is the dominant solvent component. 
 All abundances are expressed in mole fraction of the initial solution.}
\begin{center}
{\small
\begin{tabular}{lllll}
\hline
Species       &  Type A                  & Type B                & Type C                 & Type D                 \\
HCN           &  $3.519 \times 10^{-7}$ & $7.199 \times 10^{-9}$ & $4.635 \times 10^{-5}$ & $4.370 \times 10^{-6}$ \\
C$_4$H$_{10}$ &  $1.462 \times 10^{-7}$ & $7.199 \times 10^{-9}$ & $1.925 \times 10^{-5}$ & $4.370 \times 10^{-6}$ \\
C$_2$H$_2$    &  $1.381 \times 10^{-7}$ & $7.199 \times 10^{-9}$ & $1.818 \times 10^{-5}$ & $4.370 \times 10^{-6}$ \\
CH$_3$CN      &  $1.191 \times 10^{-8}$ & $7.199 \times 10^{-9}$ & $1.569 \times 10^{-6}$ & $4.370 \times 10^{-6}$ \\
CO$_2$        &  $3.519 \times 10^{-9}$ & $7.199 \times 10^{-9}$ & $4.635 \times 10^{-7}$ & $4.370 \times 10^{-6}$ \\
C$_6$H$_6$    &  $2.707 \times 10^{-9}$ & $7.199 \times 10^{-9}$ & $3.565 \times 10^{-7}$ & $4.370 \times 10^{-6}$ \\
\hline
\end{tabular}
}
\end{center}
\label{mixtureTYPES}
\end{table}
%
%
  Our model, based on PC-SAFT, has been mainly validated using data from solid organic dissolution where the solvent (methane or ethane) 
comprised the major components. However, our model has not been formally validated for very high concentration of solutes, \textit{i.e.} 
for circumstances where the sum of their mole fractions is larger than $\sim 50$\%. We have then chosen to stop the evaporation algorithm 
when $\sum_{k ({\rm solutes})} x_{k} > 0.5$. In practice, this criterion has been satisfied at the very end of the evaporation, \textit{i.e.} 
when the ratio of the remaining volume and of the initial volume was approximately between $10^{-4}$ and $10^{-8}$, depending on the 
particular composition adopted at the starting time. In fact, along the evaporation process the mole fraction of the solvent is nearly constant. 
Roughly, when a number of moles $\Delta n$ are removed from the solvent by evaporation, a similar amount $\Delta n$ of organics saturate and
settled to the lakebed.\\
       At the very end of the evaporation (\textit{i.e.} during the last time-step), the remaining liquid is assumed to evaporate, and 
the total amount of still-dissolved compounds are deposited on the lakebed. In our model, when this evaporite final layer is composed of several species, 
they are assumed to be perfectly mixed. 
%
%
\subsection{\label{maxdepth}The Maximum Thickness of Evaporite: a First Approach}
   Solution theory enables our model to estimate the allowed maximum thickness. For a given volume of liquid, the concentration at saturation 
of the considered compound gives the maximum quantity of matter that can be dissolved. Then, if we assume that all evaporite components are 
initially present in the solution at their saturation abundances, then the algorithm implemented in our model provides the thickness 
of the corresponding deposition. Table~\ref{saturationSOLUB} shows the resulting depths, denoted $e_{100}$, corresponding to an initial 
height of solvent of $100$ m. Unsurprisingly, ethane leads to greater final thicknesses because this molecule is a much better solvent than methane 
for the expected hydrocarbon solutes available to a Titan lake system. The resulting total thickness are respectively $0.324$ and $13.65$ meters,
for CH$_4$ and C$_2$H$_6$. The question remains, however, how easy it is to meet the conditions for the simultaneous saturations of all species. 
In other words, is the atmospheric photochemistry able to provide large enough quantities of organics to allow saturation in the lakes?\\
%
\begin{table}[htbp]
\caption[]{Radius (in m) of possible idealized ``catchment basins'' of solid organics required to ensure the 
           saturation of a given species in a column of solvent (CH$_4$ or C$_2$H$_6$) which height is $H$, and with a 
           cross-section of $1$ m$^{2}$ (corresponding to a disk with a radius of $0.56$ m). 
           All the computations were performed assuming an atmospheric precipitation timespan of one Titan year.}
\begin{center}
{\small
\begin{tabular}{lllll}
\hline
              &  CH$_4$       &  CH$_4$        & CH$_4$       & C$_2$H$_6$   \\
Name          &  $H= 100$ m   &  $H= 10$ m     & $H= 1$ m     & $H= 10$ m    \\
\hline
HCN           &  $15.18$      &  $4.799$       &  $1.518$     & $48.41$      \\
C$_4$H$_{10}$ &  $1621$       &  $512.5$       &  $162.06$    & $3334$       \\
C$_2$H$_2$    &  $284.2$      &  $89.88$       &  $28.42$     & $259.0$      \\
CH$_3$CN      &  $28.72$      &   $9.081$      &  $2.872$     & $167.1$      \\
CO$_2$        &  $400.2$      &  $126.5$       &  $40.02$     & $148.7$      \\
C$_6$H$_6$    &  $24.75$      &  $7.826$       &  $2.475$     & $441.9$      \\
\hline
\end{tabular}
}
\end{center}
\label{precip_sat}
\end{table}
%
%
    In Table~\ref{precip_sat}, we estimated the dimensions of the catchment basins required to dissolve
enough solid organics for the lake to reach saturation, for each investigated solute \REVsec{produced at the rates
computed by \cite{lavvas_etal_2008a,lavvas_etal_2008b} and accumulated during one Titan year}. In this scenario, solid organics fall from the atmosphere 
(in the form of snows or hail) and would be washed into the lake with rainfall runoff flowing to the local topographic minimum, the lake.
For idealized disk-shaped basins, we list the dimensions as a function of lake radius in Table~\ref{precip_sat}
for either CH$_4$ or C$_2$H$_6$ playing the role of solvent. Different liquid depths $H$ of the central are also considered. The lake 
itself is supposed to cover an area of 1 m$^2$. The computed radii correspond then to the catchment basin size required to get the 
saturation in a volume of $H \times 1$ (m$^3$) of liquid hydrocarbons.\\
  HCN and C$_4$H$_{10}$, the solutes with respectively the smallest and largest mole fractions at saturation,
need the smallest and the largest collecting area according to PC-SAFT calculation (see Table~\ref{saturationSOLUB}). More interestingly, 
these calculations indicate that butane is so soluble that reaching saturation requires an unreasonably large basin. In the most favorable 
case where methane is the solvent, for a diameter of $\sim 20$ km (typical of some northern lakes \cite[see Fig. 1 of][]{barnes_etal_2011}),
and the initial depth fixed to $1$ m, the drainage basin for butane must have a radius larger than the radius 
of Titan itself. In contrast, the saturation concentration of HCN is reached in a similar lake when this compound is drained over an area with
a radius around $26$ km while the assumed lake has a radius of $10$ km.\\

 Given the numbers reported in Table~\ref{precip_sat}, we can safely conclude that all of the solutes considered in this study
cannot be simultaneously at saturation in the initial state (\textit{i.e}. before a significant evaporation episode) for a given lake. 
In addition, if the fraction of surface covered by evaporite in polar regions can be as high as $\sim 10$\% 
\citep[see for instance Fig. 3 of][]{macKenzie_etal_2014}, the average catchment area can only have a radius of $1.78$ m for a central lake 
of $1$ m$^2$ (a disk with a radius of $1.78$ m has an area of $\sim 10$ m$^{2}$). This value is lower than the majority of radii given 
in Table~\ref{precip_sat}, indicating the improbability of lakes being saturated in their initial state. 
Thus we can firmly state that the thicknesses mentioned above, $0.324$ and $13.65$ meters, are largely an overestimate if we impose a 
timescale of one Titan year. However, in Sect.~\ref{poss1Dstruc}, we will discuss a mechanism for repeated dissolution-evaporation-deposition
that could overcome these limitations.
%
\subsection{\label{poss1Dstruc}The Possible 1D Structure of Evaporites}
\begin{figure*}[!t]
\begin{center}
\includegraphics[width=13 cm, angle= -90]{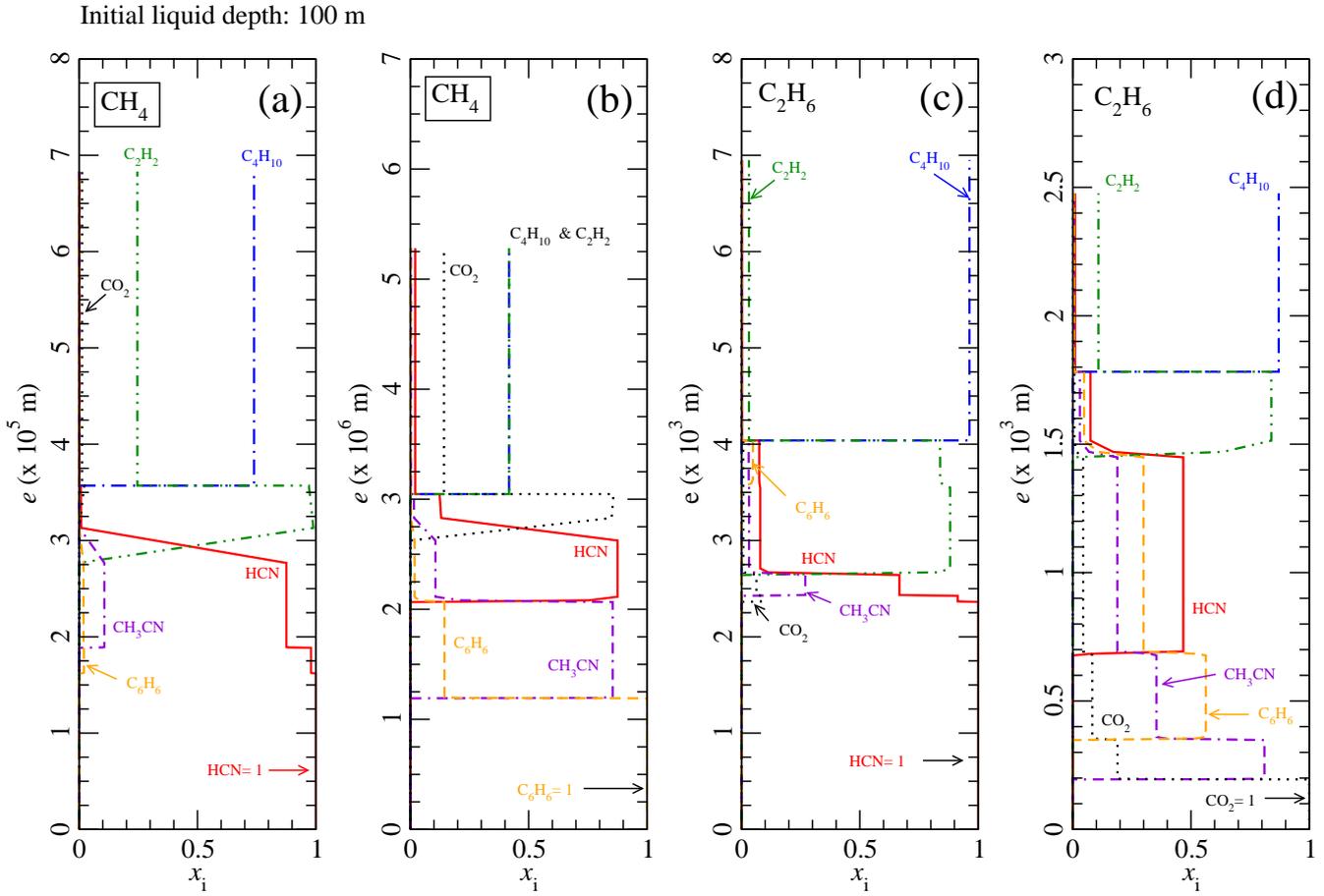}
\caption[]{\label{struct1D}1D structure of evaporite layers, computed from an initial liquid depth of 100 m. \REVsec{The
          $y$-axis represents the height $e$ of the resulting deposit over the non-soluble substrate. The $x$-axis
          shows the mole fractions $x_i$ of the differing species at a given height.}
          Two solvents are considered: methane (panels (a) and (b)), and ethane (panels (c) and (d)).
          In simulations reported in panels (a) and (c), the assumed initial mixture of solutes has
          a mole fraction of HCN (the most abundant species in atmospheric precipitation) set to
          its value at saturation in the considered solvent, and other concentration are derived by a
          scaling to atmospheric production rates. The results plotted in panels (b) and (d), have
          been obtained by adopting uniform solute concentrations, fixed to the lowest mole fraction
          at saturation.}
\end{center}
\end{figure*}

    The structures of evaporitic deposit left at a lakebed after the entire evaporation of an assumed $100$ m-high column of liquid
have been computed. The results shown in Fig.~\ref{struct1D} can be rescaled for any other initial liquid height. Two solvents have 
been employed: pure methane and pure ethane. In panel (a) of Fig.~\ref{struct1D} the initial mixtures of dissolved organics are set 
by fixing the concentration of the most abundant species (\textit{i.e}. HCN) to its value at saturation (\textit{i.e.} 
$3.52 \times 10^{-9}$, see Table~\ref{saturationSOLUB}) while the abundances of other compounds are 
derived by scaling to the atmospheric production rates; the result is the ``type A''
mixture in Table~\ref{mixtureTYPES}. As a consequence, the initial total mole fraction of solutes reaches only $6.54 \times 10^{-7}$, 
a value that can be converted to a deposit thickness of a few tens of micrometers ($6.83 \times 10^{-5}$ m).\\
   As already emphasized in PAP1, the most soluble species, \textit{i.e.} C$_4$H$_{10}$ and C$_2$H$_2$, remain dissolved until the very 
end of the evaporation process. Thus, these species dominate the final, top layer of the deposit. We noticed that the last droplet of 
solution to be evaporated contains the entire amount of dissolved C$_4$H$_{10}$ and C$_2$H$_2$, around $50$\% of the volume of dissolved 
material. This is due to the small total amount of solute and the high solubilities of C$_4$H$_{10}$ and C$_2$H$_2$ and explains the final 
vertical parts of the curves in the top of Fig.~\ref{struct1D}(a). Alternatively, the least soluble compound\JWB{s} (dominantly, for this mixture, HCN) 
are buried below the C$_4$H$_{10}$- and C$_2$H$_2$-enriched top layer.\\
   The treatment of the ``last droplet to be evaporated'' is worth special attention. Indeed, our model has not been validated 
in situations where the solvent becomes a minor species of the ``solution''. Thus, we have simply adopted a principle of a well-mixed last layer, 
which is reflected in the top vertical parts of curves in Fig.~\ref{struct1D}(a) (and other panels). We therefore ignore possible 
segregation effects (presently unknown) that could occur during the late stages of evaporation. However, we are aware that different 
species can precipitate under different crystallographic phases leading to an inhomogeneous mixture. This aspect will be the subject of 
a point of discussion further on in the paper.\\

  Although the ``type A'' mixture is more realistic, we also use an uniform initial distribution of solutes (``type B'' mixture displayed 
in Table~\ref{mixtureTYPES}). With this mixture, we eliminate the effect of the initial mixing ratios on the evaporite structure. 
HCN is no longer the solely dominant buried species; CH$_{3}$CN and C$_6$H$_{6}$ play a prominent role in this scenario. While butane and 
acetylene are still the major compounds of the external layer, carbon dioxide appears to reach abundances around $14$\%.\\
   We also examine pure ethane as the solvent in the ``type C'' and ``type D'' mixtures of Table~\ref{mixtureTYPES}.  Mixture ``C'' has 
an initial mixture corresponding to the most abundant species in precipitation (\textit{i.e.} HCN) taken at its saturation. The initial 
solution for mixture ``D'' is similar to ``type B'' in that the uniform initial mole fractions are used: $4.37 \times 10^{-6}$, the 
lowest concentration at saturation of our set of solutes, \textit{i.e.} that of CO$_2$. The results for these mixtures are displayed 
in Fig.~\ref{struct1D}.(c)  and Fig.~\ref{struct1D}.(d), respectively, where the scale factor attached to the $y$-axis is $10^{-3}$, 
meaning a depth of the order of a few millimeters. Solutions with methane (types ``A'' and ``B'') produce only micron-deep layers.\\
%
%
%
%
%
%
  In the case where initial abundances are scaled to respective precipitation rates (``type C'', Fig.~\ref{struct1D}.c), the 
structure can be generally described as a layer of less solubles compounds (\textit{i.e.} the nitriles CH$_3$CN and HCN) topped 
by a layer of butane. Comparison to Fig.~\ref{struct1D}.(a) shows unambiguously that the ethane-based mixture favors butane 
while the methane-based mixture favors acetylene in the surface layer. When uniform initial fractions are assumed (see mixture 
``type D'' in Table~\ref{mixtureTYPES}), butane remains the most abundant species at the surface but leaves some space for C$_2$H$_2$. 
Not surprisingly, the interior structure is more complex with a non-negligible role of carbon dioxide and benzene.\\

  If we focus on the external layer, these computations employing our new model based on PC-SAFT confirm the tendencies found in PAP1. 
However, we find here that the total thickness of the deposits is on the order of a few microns for a methane-rich solvent and a several 
millimeters for an ethane-rich solvent. In addition, when initial abundances of solutes are scaled to precipitation rates for either solvent, 
HCN seems to be the dominant buried species.\\

%
  In order to assess the possible influence of dissolved N$_2$, we introduce an amount of nitrogen fixed at $10$\% of the 
current quantity of either CH$_4$ or C$_2$H$_6$. This mixing ratio seems realistic according to the current literature
\citep[]{cordier_etal_2009,cordier_etal_2013a,glein_shock_2013,tan_etal_2013,luspay_kuti_etal_2015}. 
  Our results do not significantly differ with the inclusion of N$_2$ in the four mixture types. 
\REVsec{For instance, the resulting sequence of a 1D model of evaporites deposit structure remains essentially unchanged when the fraction of N$_2$ 
in the solvent (mainly composed by CH$_4$) is increased from $0.00$ to $0.20$. The most important change is a decrease of $\sim 15$\% of the abundance of 
C$_2$H$_2$ in the top layer while an increase of the C$_4$H$_{10}$ mole fraction of $\sim 7$\% is found.
This general low sensitivity to dissolved nitrogen abundances is not a surprise} because the 
PC-SAFT interaction parameters $k_{ij}$ between N$_2$ and the introduced solvent were set to zero due to the lack of relevant data. 
We emphasize, nevertheless, that except in the case of very strong interaction between these species, the role of N$_2$ should be of 
small importance, due to its relatively small abundance in the solutions.\\

\begin{figure}[!t]
\begin{center}
\includegraphics[width=7.2 cm]{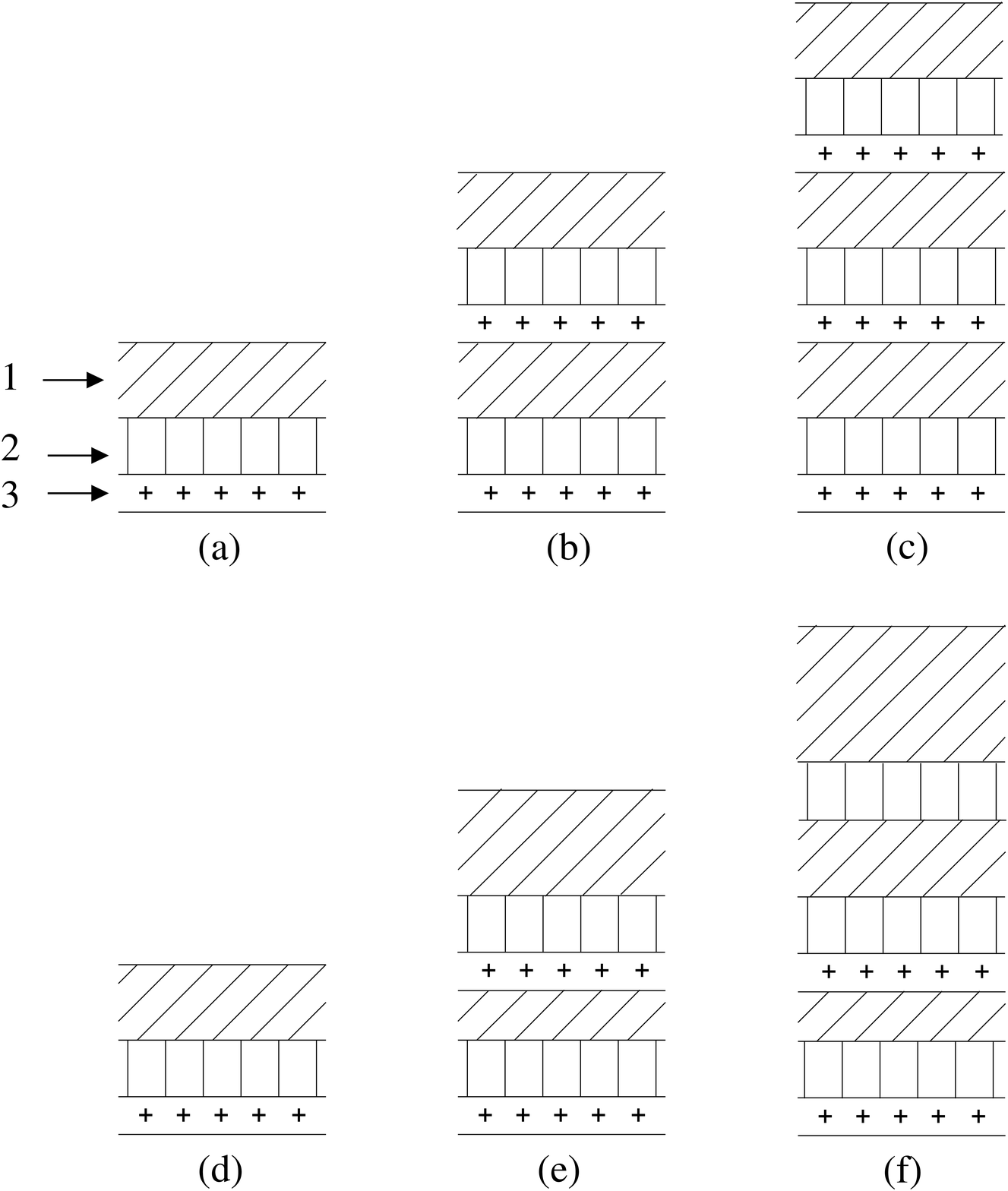}
\end{center}
\caption[]{\label{sandwich}Sketches of two scenarios of evaporites formation: three species (labeled 1, 2 and 3) are assumed for the sake of clarity. 
           Both rows illustrate the evolution of a lakebed's evaporite layers over a period of three drying and evaporating cycles.
           The series (a), (b), and (c) corresponds to the accumulation of layers without any dissolution of the subtrates deposited 
           during the past years. 
           At the bottom, the series (d), (e) and (f) presents schematically the  evolution of 
           evaporite layers if redissolution of previously laid-down layers is allowed to occur (in this example, between (e) and (f)
           only the two previous external layers were redissolved).}
\end{figure}
%
   Up to this point, all the simulated deposits were implicitly formed over an insoluble substrate. This approach is particularly 
relevant if this substrate is made of water ice and/or long chain hydrocarbons. Nevertheless, one can well imagine lakebeds successively 
flooding and drying year after year. If the atmospheric products are still the same in nature and quantities from one year to the next, 
we can expect an accumulation of evaporite at the bottom of these lakebeds. Additionally, liquid flowing into the system can re-dissolve, 
at least partially, the strata formed in previous cycles. We represent the end-member scenario where liquid runoff does not dissolve 
previously-formed evaporite layers in panels (a), (b) and (c) in Fig.~\ref{sandwich}, which could happen if the liquid runoff is 
too fast over the deposits. 
  However, for liquid entering the system that is in contact with the deposits long enough
to dissolve the evaporite \REVsec{\citep[experiments by][give credence to this assumption]{malaska_hodyss_2014}}, 
panels (d), (e) and (f) depict the evolution of evaporite layers.  Around $\sim 50$\% of the top layer formed during the previous year 
is brought into solution. As shown above, the most soluble species build the external layer of evaporite (deposited during some
previous epochs). They are then the first ``re-dissolved'' material such that the current liquid solution becomes \REVsec{more enriched}
in the most soluble compounds.
 This leads to a secondary surface layer over-enriched in butane and/or acetylene (see panel (e) in Fig.~\ref{sandwich}).
%
This process can be repeated from year to year, yielding to a very thick layer of the most soluble species (Fig.~\ref{sandwich}.f)\REVsec{, 
though the process might be limited by the saturation points of the solutes and the quantity of solvent running off}.\\
%
%
%
%
\begin{figure}[!t]
\begin{center}
\includegraphics[width=6 cm]{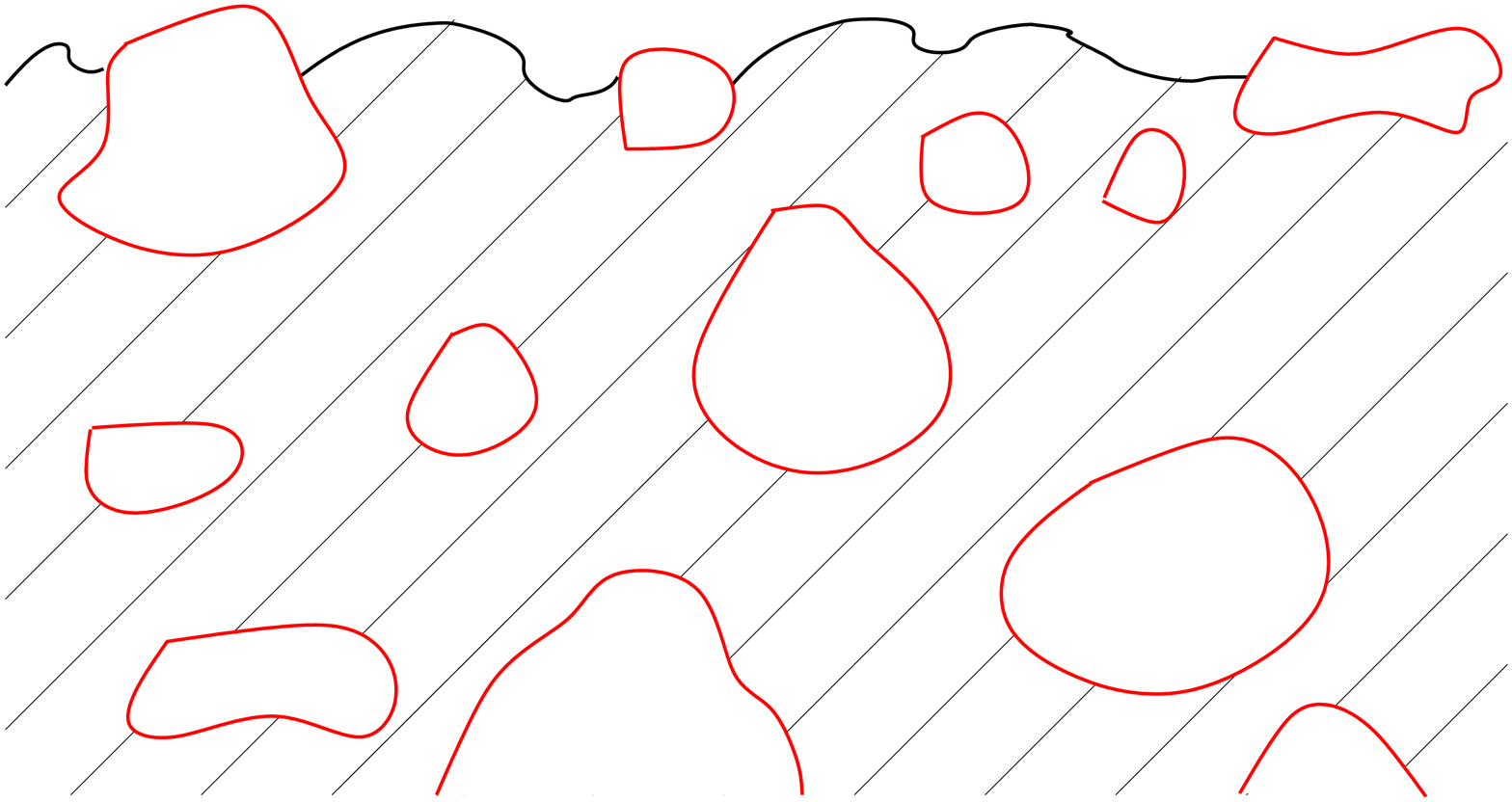}
\end{center}
\caption[]{\label{toplayerstruct}Scheme of the plausible top layer structure of Titan's evaporite. The hatched area corresponds to
           a matrix made of the most abundant compound \REVsec{(forming a first macroscopic crystallographic phase)}, 
           blank zones are composed by the less abundant, \REVsec{this corresponds to a second macroscopic phase. These heterogeneities have
           an effect on RADAR backscatter only if their scale is at least comparable to RADAR wavelength}. }
\end{figure}
%
      Alternatively, if a much larger quantity of material is dissolved, for instance if all the deposited organics during the past year 
are redissolved, then, the new layers deposited in the next evaporation event are much thicker. In all cases, the dissolution of the 
previously formed layer increases the thickness of the external layers. This effect, however, is limited by the concentration at 
saturation and the average annual precipitation of liquid methane/ethane.\\
%
%
%
    \REVsec{Regardless} \REVfirst{of} the initial composition scenario, the surface layer of evaporite could be composed by a mixture of 
butane and acetylene. In addition, this top layer appears to be relatively thick compared to the depth of the whole deposit. With multi-annual 
repetition of the dissolution-evaporation process, the top butane-acetylene rich layer tends to grow in thickness. In Fig.~\ref{struct1D}, 
the uniform mole fractions of C$_2$H$_2$ and C$_4$H$_{10}$ are only relevant on average. Indeed, it is extremely likely that these 
two species precipitate separately, each one in its own crystallographic phase. An homogeneous phase of C$_2$H$_2$-C$_4$H$_{10}$ 
would imply the existence of something ressembling an acetylene-butane ``co-crystal''
yet unknown but similar to what has been observed by \cite{vu_etal_2014} for ethane and benzene.
In addition, even if this kind of 
system exists, a perfectly homogeneous layer would require relative abundances of C$_2$H$_2$ and C$_4$H$_{10}$ in agreement with the 
allowed stoichiometry of the ``co-crystal''. Thus, if butane and acetylene are present, the existence of a biphasic system seems more
likely, and the structure of the top layer of evaporite could be similar to that depicted in Fig~\ref{toplayerstruct}, \REVsec{although
the scales of heterogeneities are unknown.}\\

   The pebbles observed at \textit{Huygens} landing site were probably formed by mechanical erosion and required relatively powerful liquid 
currents to flow in Titan's rivers \citep{tomasko_etal_2005}. In contrast, evaporation in small lakes and ponds is a more gentle process but would 
also leave some irregularities \REVsec{like pebbles, evaporitic polygonal crusts, macroscopic crystals, that could produce high radar 
brightness as it has been also speculated in the case of channels observed by the Cassini RADAR \citep[][]{legall_etal_2010}. Indeed,}
   on Earth, Devil's Golf Course (Death Valley, California) or Lucero Lake (White Sands National Monument, New Mexico)
offer examples of evaporitic formations that show a several tenths of centimeters in size rugosity. Although these structures were
largely due to erosion, there is no reason to ensure that this situation does not occur on Titan. \REVsec{Hence, the RADAR brightness of evaporites
\citep[see][ Sect. 3]{barnes_etal_2011} could be explained by processes occuring during either the lacustrine basin formation or the formation
of evaporite deposits.}\\
  Unfortunately, we did not find frequency-dependent permittivity for solid butane and acetylene in the literature. However, since 
C$_2$H$_2$ and C$_4$H$_{10}$ are both non-polar molecules, their polarizability provides the main contribution to the permittivity 
of their solid forms. We obtained the static permittivity $\epsilon_{r,0}$ at $195$ K for both species ($\epsilon_{r,0}$(C$_4$H$_{10}$)$= 1.942$ 
and $\epsilon_{r,0}$(C$_2$H$_2$)$= 2.4841$) from \textit{Handbook} \citep{handbook74th}. The difference in these values suggests substantially 
different permittivities in the microwave domain of the Cassini RADAR. We have shown that the thickness of evaporite can be as large as 
several tens of centimeters or even several meters, much larger than the wavelength of the Cassini RADAR (\textit{i.e.} $2.18$ cm)
and therefore potentially affecting the RADAR signal.
\REVsec{The observed RADAR brightness can be also caused by a layering but probably more likely by heterogeneities as depicted 
in Fig.~\ref{toplayerstruct}.}\\
%
%
   \REVsec{In summary, the RADAR brightness at the evaporite \citep[][]{barnes_etal_2011}, if not produced by centimeter-sized surface roughness,  
could be also caused by heterogeneities within the top layer produced by the existence of at least two crystallographic phases. 
The subsurface horizons generated by stratification of evaporite (see Fig.~\ref{struct1D}) could also contribute to the effect, but the formation
of plane interfaces between layers could be more difficult.}
%
%
%
\begin{table}[t!]
\caption[]{Lake shore slopes derived from data published by \cite{hayes_etal_2008} (Fig.~3, panel c). In this paper, the positions 
           of the lakes are indicated by their respective abscissa along the RADAR track, \textit{i.e.} ``shore 1'' is the one first 
           crossed by the track. The central part of the ``lake 1'' bed has not been taken into consideration because of the lack of data.
           ``Lake 1'' and ``lake 2'' correspond respectively to features around $\sim 140$ and $\sim 320$ km along RADAR track in Fig.~3
           of \cite{hayes_etal_2008}.}
\begin{center}
{\small
\begin{tabular}{lccc}
\hline
Object                  & Shore 1 Slope      &  Central region of the bed   & Shore 2 Slope   \\
                        & (in degrees)       &  (in degrees)                & (in degrees)    \\
\hline
Lake 1                  & 1.812              &   ---                        & 1.773           \\
Lake 2                  & 4.445              &   0.420                      & 3.180           \\ 
\hline
\end{tabular}
}
\end{center}
\label{shoreslopes}
\end{table}
%
\section{\label{2Dmodel}Evaporite Deposits Structure: a 2D Model}

    \cite{barnes_etal_2009a,barnes_etal_2011} and \cite{macKenzie_etal_2014} observed 
evaporite deposits along the periphery of lakebeds. The signal at 5 $\mu$m shows a gradient that could be explained by changes in chemical 
composition and/or thickness of the deposited organic material. On Earth, the combination of drought and increased water 
demand has produced significant drops in water levels of the well-known reservoirs Lake Mead and lake Powell. Consequently, ``bathtub rings''
have appeared along the shores of these lakes. These structures, mainly made of calcium carbonate, are observable in pictures taken from space 
\cite[see for instance][Fig. 7]{barnes_etal_2009a}. \REVsec{Similarly-formed ``bathtub rings'' around lakes on Titan are probably more complex due to 
the variety and different properties of the organic compounds involved. Consequently, these particular formations could be unique through the solar system}.
%
   \REVsec{Thus, a better understanding of evaporite formation is desirable}.

\begin{figure}[!t]
\begin{center}
\includegraphics[width=8.0 cm]{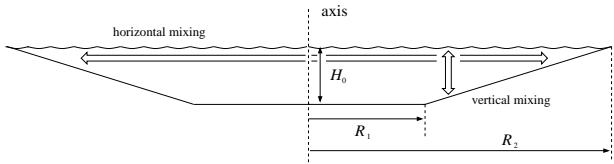}
\end{center}
\caption[]{\label{lakebedscheme} Schematic cross-section of our idealized lakebed. 
           We denote $H_{0}$ the initial liquid depth, $R_{1}$ and  $R_{2}$ are defined in the text.}
\end{figure}

   For the sake of simplicity, we have adopted axisymmetric topography, as sketched out in Fig.~\ref{lakebedscheme}. The bottom of 
the lakebed is represented
%
%
%
    by a disk-shaped flat terrain of radius $R_{1}$. This zone is surrounded by sloping ground 
that extends between the radii $R_{1}$ and $R_{2} > R_{1}$. The value $R_{2}$ corresponds to the area covered by a volume with 
initial liquid depth $H_{0}$. \cite{mastrogiuseppe_etal_2014} have performed bathymetric measurements along a RADAR track acquired 
during a nadir-looking altimetry flyby above Ligeia Mare. This sea is much larger than the class of lakes we are interested in. 
%
%
%
\begin{figure*}[!t]
\begin{center}
\includegraphics[width=13 cm]{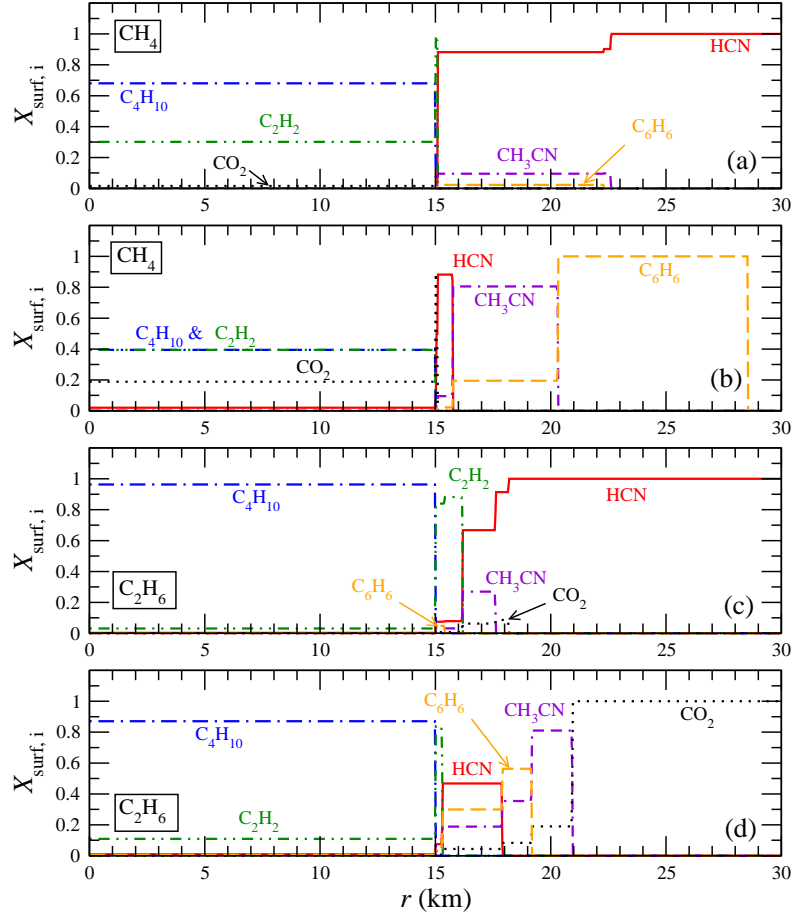}
\end{center}
\caption[]{\label{surfacecompo}Surface composition (in mole fraction) of a model lakebed with topography depicted in
Fig.~\ref{lakebedscheme}. The initial mixtures are those adopted for simulations reported in Fig.~\ref{struct1D}, \textit{i.e.} in
panels (a) and (b) the solvent is pure CH$_4$ while in panels (c) and (d) it is ethane. Two solute compositions are considered:
the initial abundances are scaled to atmospheric precipitation rates (with HCN initially at saturation), 
corresponding to panels (a) and (c), and uniform abundances, as in panels (b) and (d). In order to be consistent with Fig.~\ref{struct1D},
species are represented by the same line style. The radius $r$ represents the distance to the axis of the modeled lake basin.}
\end{figure*}
%
We note, however, that the shallow slope of the seabed revealed by the global bathymetric profile 
\citep[see Fig. 4 of][]{mastrogiuseppe_etal_2014} generally agrees with the slope scheme we adopt here. 
Moreover, 
our results do not depend on the precise slope but rather on the exact shape of the shore terrain.
   In our baseline scenario, continuous evaporation removes liquid from the system while solid compounds are deposited in the bed. 
Parts of the lakebed deposits that are no longer immersed maintain their structure and composition until the end of the process. 
It is implicitly assumed that the kinetics of both precipitation and sedimentation are much faster than the kinetics of evaporation. 
Parts of the lakebed that are still submerged get covered by a growing layer of solid organics that \REVfirst{stratifies} gradually.\\
   The properties and global characteristics of liquids mixing in Titan's lakes remain relatively unknown. 
Different physical processes can contribute to this mixing: vertical convection \citep{tokano_2009}, tidal effects 
or global circulation \citep{tokano_etal_2014}. In our approach, we chose to ignore the possible details of 
this mixing and instead consider two extreme cases: (1) only an efficient vertical mixing occurs 
and (2) a combined horizontal-vertical mixing scheme that ensures chemical homogeneity of the entire lake (see Fig.~\ref{lakebedscheme}).\\
   In case (1), the whole quantity of dissolved solutes contained in the initial column of liquid above a given 
point of the lake is precipitated on the bed following the behavior reported in our 1D model study 
(see Sec.~\ref{1Dmodel}). In such a case, the resulting composition of the evaporite deposition will show an uniform surface 
composition: only the depth will vary from one point to another. Locally the thickness of evaporite layers scales to the initial height of the 
local liquid column. No bathtub ring structures are expected in case (1).
Thus, observations of a Titan lakebed lacking evaporite rings
could be understood as the mark of a weakly efficient horizontal mixing in the lake and/or 
evidence for an unsaturated solution.\\
   In case (2), the time-dependent compositions and thickness of the deposits are easily computed by fixing the sizes of the lake to typical 
values, \textit{i.e.} $R_{1}= 15$ km, $R_{2}= 30$ km \citep[see for instance Fig. 1 panel C of][]{barnes_etal_2011} and an initial depth 
of $H_{0}= 600$ m. This implies a shore slope of $\sim 2.3^{\rm o}$, similar to those reported in Table~\ref{shoreslopes}. The 
computation algorithm is divided in two main steps. First, for the corresponding total volume $V_{tot}=\pi H_{0} (R_{1}^{2}+R_{2}^{2}+R_{1}R_{2})/3$, 
our thermodynamic model is employed to estimate the quantities of precipitated matter during each time step. The resulting outputs are then
applied to the chosen particular geometry such that at each timestep, the total amount of over-saturated species is uniformly 
distributed over the immersed part of the bed. These operations are repeated until the solvent is exhausted.\\

   The results are summarized in Fig.~\ref{surfacecompo}. The 
mixture types used for the 1D model are also used here. The solvents are pure methane (Fig.~\ref{surfacecompo} panels (a) and (b)) and 
pure ethane (Fig.~\ref{surfacecompo} panels (c) and (d)). For each solvent, two initial solutes compositions are considered: 
either the abundances are scaled to atmospheric precipitation rates (Fig.~\ref{surfacecompo} panels (a) and (c)), or the initial mole 
fractions are all fixed to the smallest saturation value (Fig.~\ref{surfacecompo} panels (b) and (d)). Unsurprisingly, the surface of 
the central part of the evaporite deposits, \textit{i.e.} that which covers the flat bottom of the basin, has a composition dominated 
by butane and acetylene. This behavior is explained, as in the 1D results, by the large solubilities of these two species, which are thus
able to remain dissolved until the very late stages of the liquid evaporation. During this last episode, the liquid stagnates above 
the bed bottom and the species, that are still dissolved, finally precipitate out. Clearly, this composition is consistent 
with the abundances of the top layers exhibited by the 1D model (see Fig.~\ref{struct1D}), although a slight difference is 
evident in the ethane solvent (Fig.~\ref{struct1D}.c) where an almost pure butane region is surrounded by an acetylene rich crown.\\

   The most external parts of the lacustrine basin are covered by a surface made of solid HCN when 
the solutes initial abundances are scaled to the atmospheric precipitation composition (see Fig.~\ref{surfacecompo} 
panels (a) and (c)). In this scenario, the most plausible for an average lake, hydrogen cyanide is the 
most abundant dissolved compound though it is very poorly soluble. When the initial composition of solutes is uniform 
(Fig.~\ref{surfacecompo} panels (b) and (d)), an external ring is made of either benzene (when CH$_4$ is 
the solvent, Fig.~\ref{surfacecompo}.b) or carbon dioxide (when C$_2$H$_6$ is the solvent, Fig.~\ref{surfacecompo}.d). These results 
indicate that the occurrence of an external HCN-rich ring would likely be caused by large initial 
content of the solution rather than a pure solubility effect. We emphasize that 
although HCN is clearly detected in the atmosphere \cite[see for instance][]{vinatier_etal_2010a,dekok_etal_2014}, its solubility in cryogenic 
solvents remains not well known, and values provided in this work are less reliable than those concerning other species, especially when 
the model outputs are compared to experimental works (see Sect.~\ref{modsolub}). \REVsec{Laboratory experiments are needed in order to 
determine the interaction parameters $k_{ij}$s related to HCN.}\\

  Finally, between the outer portion (\textit{i.e.} $r \gtrsim 22$ km) and the central area (\textit{i.e.} $r \lesssim 15$ km) there
lies a transitional zone that exhibits a chemically complex surface. Whatever the initial assumed composition, the resulting surface 
composition of evaporites appears to be ``trimodal": a C$_4$H$_{10}$-C$_2$H$_2$ central region is bordered by a chemically complex narrow ring 
 which is itself surrounded by an extended region where HCN is the dominant species (if the adopted atmospheric precipitations are 
 representative of the actual weather conditions in Titan's troposphere). These conditions could either bring solutes to the lake 
directly from atmospheric fallout or wash the surrounding terrains and dissolve pre-existent solids that could cover these areas.\\
  We underline that this chemically ``trimodal'' surface composition does not depend on the actual slope of the lacustrine basin shore. 
Indeed, a shore with a more gentle slope will be covered by thinner depositions, but will show more extended ``bathtub rings'', the 
aspect ratio being preserved. Narrower rings will be caused by steeper shores, again maintaining the aspect ratio. In terms of deposit 
thickness, low slopes correspond to shallow evaporite layers, whereas steep shores will exhibit thick strata.\\

   The spectroscopic observation of all species involved in this study, is beyond the capabilities of an instrument like VIMS. 
However, the predicted ``trimodal'' surface composition of evaporite layers could be tentatively detected  by VIMS if the spatial 
resolution is high enough (for instance $\lesssim 5$ km/pixel).  The data spanning Ontario Lacus' evaporite-covered shorelines,
analyzed by \cite{barnes_etal_2009a}, have a high spatial resolution, as good as $330$ m/pixel. These coastal features seem to be 
``bimodal" with two distinct zones \citep[see ][Fig. 4]{barnes_etal_2009a}. After excluding several hypotheses (freezing, continental 
shelf, etc.), \cite{barnes_etal_2009a} proposed that the inner ring could be an intertidal zone showing exposed lake-bottom sediments. 
The external ring appears to have a low water-ice content, leading \cite{barnes_etal_2009a} to propose that it consists of
fine-grained condensate, resulting of the evaporation of the liquid. These observations are consistent 
with our ``trimodal" evaporation deposition. Indeed, in the case of Ontario Lacus, the liquid could still contain a large amount of 
solutes since it seems to be rich in ethane \citep{brown_etal_2008,luspay_kuti_etal_2015}, 
a much more efficient solvent than methane. In our simulations, the central part of the deposit (see Fig;~\ref{surfacecompo}) is built 
up during the last stages of the evaporation, when butane and/or acetylene crystallize. Hence, the distinct two zones observed by 
\cite{barnes_etal_2009a} could correspond to two chemically different evaporite deposits: 
``unit 3" (the most external in Barnes \textit{et al}. denomination) is perhaps composed of HCN while ``unit 2'' could be a ``chemically complex''
deposit. This interpretation is not mutually exclusive with that of an intertidal zone.\\

   Also using VIMS data, \cite{moriconi_etal_2010} have tentatively detected organic species within the rings observed 
around Ontario Lacus. They used the Spectral Angle Mapper technique to compare pixel spectra to the reference spectra of compounds 
of interest (C$_2$H$_6$, CH$_4$, C$_4$H$_{10}$, HCN, C$_3$H$_8$, C$_2$H$_2$ and C$_6$H$_6$).
However, the definitive identification of surface compounds on Titan remains a matter of debate given the few opportunities left to 
see the surface with a reduced atmospheric contribution and an increased Signal-to-Noise Ratio with VIMS.\\

   Nonetheless, the infrared data analyzed by \cite{moriconi_etal_2010} appear to be compatible with the presence of C$_4$H$_{10}$, 
C$_2$H$_2$ and HCN within the area called ``the ridge'' by the authors (equivalent to unit 3 of the study of Barnes {\it et al}., 2009). 
The possible detection of species in liquid state under Titan's 
ground conditions, \textit{i.e.} C$_2$H$_6$, CH$_4$ and C$_3$H$_8$, can be explained by either sediments still soaked in the corresponding 
liquid or an altimetric profile not as simple as that depicted in Fig.~\ref{lakebedscheme}. Indeed, citing \cite{lorenz_etal_2009}; 
\cite{moriconi_etal_2010} noted that the ``ridge'' could have a non-uniform elevation \citep[see also Fig. 6 in][]{cornet_etal_2012b}. 
A lake shore altimetric profile with a changing slope could lead to a solid crystallization sequence within sporadic pools that could exist in the zone. 

\begin{figure}[!t]
\begin{center}
\includegraphics[width=7 cm, angle=-90]{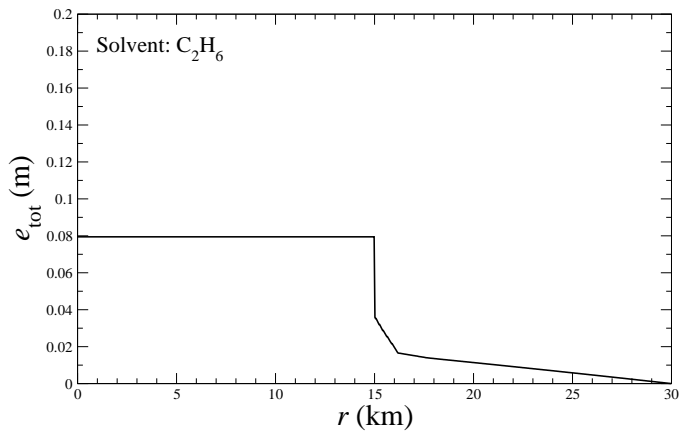}
\end{center}
\caption[]{\label{evaporitedepth}The total thickness $e_{tot}$ (in meter) of the resulting evaporite layer that appears after
           evaporation of the solvent. The adopted lakebed topography is described in Fig.~\ref{lakebedscheme}, $r$ (km) represents
           the typical radius of a lake initially containing solutes with abundances scaled to atmospheric precipitation
           rates (this model corresponds to what it is depicted by Fig.~\ref{surfacecompo}(c)).}
\end{figure}

  Globally, the findings of \cite{moriconi_etal_2010} are in agreements with our predictions where C$_4$H$_{10}$, C$_{2}$H$_{2}$ and 
HCN appears to be the most abundant species at evaporites surface. Fig.~\ref{surfacecompo} shows that C$_6$H$_6$ has a very discrete 
presence in scenarios where the initial compositions are scaled to atmospheric abundances (panels (a) and (c) in Fig.~\ref{surfacecompo}). 
Therefore, the non-detection of benzene can be interpreted as evidence that the composition of the atmospheric precipitation
is similar to that computed by \cite{lavvas_etal_2008a,lavvas_etal_2008b}, as proposed in the interpretation of \cite{moriconi_etal_2010}.
\REVsec{However, in their surface mapping of a $5.05-\mu$m spectral feature on Titan's surface, \cite{clark_etal_2010} shown that benzene could be
present in a circular geological pattern (see Fig. 17.B of their study), which they interpreted as a dry lakebed. Unfortunately, 
this feature has not been imaged by the RADAR in order to confirm its exact geological nature, and mainly whether it pertains to 
the class of possible lakebeds or not.}\\
   Our model also predicts evaporite layer thickness. In Fig.~\ref{evaporitedepth}, we have plotted the total thickness of the evaporite
deposition in the scenario where solute abundances are scaled to production rate of solids (see Fig.~\ref{surfacecompo})
and using an ethane solvent. The central plateau is explained by the final deposition of C$_4$H$_{10}$, in agreement with the results of our
1D models (Fig.~\ref{struct1D}.c). The change in slope observed between $r= 15$ km and $16$ km is a consequence of the 
sudden saturation of acetylene (see Fig.~\ref{surfacecompo}.c).\\
   Fig.~\ref{evaporitedepth} is typical of the thickness distribution of evaporite deposits after one sequence of dissolution-evaporation. 
In Titan weather conditions, successive evaporation and flooding episodes can occur and thus drive thicker deposits via redissolution/precipitation 
mechanisms similar to those already discussed in Sec.~\ref{maxdepth}.
%
%

\section{\label{discuss}Discussion}
%
%
%
    Throughout this article we have assumed that the main process that removes a
solvent is evaporation. Several authors have discussed the possibility of fluids percolation within some porous regolith or terrain 
\citep{hayes_etal_2008,choukroun_sotin_2012,macKenzie_etal_2014}. This process could be efficient only in the case where solid particles would not 
fill the regolith pores. A situation where the compounds belonging to the solvent could percolate, leaving behind initially dissolved species, 
seems particularly unrealistic since
the pores would have very specific (and still unknown) properties; in addition, deposited layers at the lake bed would have to remain
permeable to solutes, irrespective of its thickness.
Nevertheless, we could wonder whether even in case of liquid percolation, the
formation of evaporitic deposition could take place. Hence, if the bulk of the liquid, \REVfirst{initially} lying in a lacustrine depression, flows
into a porous geologic formation instead of evaporate; at the end it will still remain a thin layer of liquid. The depth of this
former layer depends on the wetting properties of the system solid substrate-liquid, namely the surface tensions liquid-vapor and
liquid-solid \citep{butt_etal_2003}, which are essentially unknown, although they do exist. A crude estimation of the possibly resulting depth of the coat of evaporite can
be made. For that purpose, we assume a wetting film of a thickness around $1$ mm. Table~\ref{saturationSOLUB} provides the thickness
equivalent to an initial depth of $100$ m of saturated liquid. From these numbers, one can easily derive the possible thickness of
the evaporite layer left after the evaporation of a final wetting film. One finds a thickness around
a few microns, mainly due to butane. Thus, even if the geological substrate is porous, the formation of a fine evaporite layer remains
possible.\\

   All of our solubilities have been computed assuming a solvent that contains three major species: CH$_4$, C$_2$H$_6$ and N$_2$. 
This was clearly a reasonable assumption; however, it does not account for the potential role of a less abundant player like propane. Although 
photochemical models predict a relative precipitation rate of C$_3$H$_8$ one order of magnitude smaller than that of ethane
\citep[][]{lavvas_etal_2008b}, propane could play a role in regions where it would be overabundant compared to its global average 
concentration. \cite{brown_etal_2008}, in their analysis of the $5$-$\mu$m window, were not able to exclude the
existence for a small amount of propane, butane and high-order alkanes in Ontario Lacus.
Thus, we have calculated the mole fraction at saturation of considered solutes in pure C$_3$H$_8$. Unfortunately, among 
the PC-SAFT interaction parameters $k_{ij}$ relevant for propane, the only one available is that of the couple (C$_3$H$_8$,C$_2$H$_2$)
\citep[taken from][]{tan_etal_2013}, for others we used $0.0$ (like in most cases involving C$_2$H$_6$ as a solvent). In general, for
a solute X, adopting $k_{ij}= 0.0$ for both (C$_2$H$_6$,X) and (C$_3$H$_8$,X), we got a maximum solubility in propane roughtly a factor
of $2$ higher than in ethane. This tendency is also found in the case of C$_2$H$_2$ where $k_{ij}$ is known. The only noticeable exception
is for CO$_2$; for this compound we use the $k_{ij} \ne 0.0$ determined for (C$_2$H$_6$,CO$_2$) 
(see Sect.~\ref{solubilitymodel}), but we have not been able to draw any satisfactory conclusions. In short, the possible presence of some amount of 
propane could somewhat enhance the solubility of solid organics without changing essentially the formation scheme of evaporites.\\

   \REVsec{It is difficult to compare our solubility estimations with previous works since the theoretical background (equilibrium with the atmosphere or not, the use
of RST or PC-SAFT, ...), the thermodynamical conditions (pressure and temperature) and the exact composition of the solvent and the solutes taken into account vary from one
publication to another. For instance \cite{dubouloz_etal_1989} and \cite{cordier_etal_2013a} consider simultaneous equilibria of the liquid with solid organics,
and the vapor phase of the atmosphere. \cite{raulin_1987} consider a temperature of $94$ K while we take $90$ K, which is probably more
realistic. In addition \cite{raulin_1987} used a version of Eq. (\ref{EQbase}) that is slightly differ from our version, but the term introduced 
in their equation is questionable (see our discussion in the \ref{append}). A summary of solubilities found in previous works can be found in \cite{cornet_etal_2015} (see their table B.5).
Making a detailed list of the explicit differences in the theoretical assumptions of each published model would not reveal any 
physical insight great enough to justify such an undertaking.
However, one can notice that they have a significant dependence on free parameters in common. The values of these parameters
are directly determined by adjustement on experimental data, as these are the best numbers currently available.
There is never, however, a guarantee of their
validity in the context of Titan. Sometimes the values are crudely estimated, like the $\delta_{ij}$ in the case of the RST, or a default value is adopted as is the case
in \cite{stevenson_etal_2015b} and in this work for nitrogen compounds (the $k_{ij}$'s are taken equal to zero for the nitrogen species). In this work, we made a 
subtantial effort to get an agreement with available experimental data by adjusting free parameters. This could explain the differences 
between some previous works \cite[][]{raulin_1987,dubouloz_etal_1989,cordier_etal_2013a}
and our results for C$_2$H$_2$, CO$_2$, and C$_6$H$_6$. We determined that a disagreement of the order of about a factor of ten remains
acceptable. Recently \cite{stevenson_etal_2015b} concentrated their experimental efforts on nitrogen compounds. As a general trend, they found HCN and CH$_3$CN less soluble than what
we found. Nonetheless, the solubility of CH$_3$CN in pure methane at $94$ K (see their Table 4) is close to our determination: both are around $10^{-8}$. Other models
provide higher values, \textit{e.g.} COSMO-RS estimates around $10^{-7}$. In the case of HCN, the output of the Stevenson {\it et al}.'s COSMO-RS model
is not too far from our value with solubilities of the order of $10^{-7}$--$10^{-8}$. The implementation of PC-SAFT by \cite{stevenson_etal_2015b} yields values
significantly smaller, with solubilities in the range of $10^{-11}$--$10^{-12}$. 
We recall that this could be explained by parameters that are dissimilar between the two models: $m$, $\sigma$
and $\epsilon$ were directly determined with laboratory data in our work (see our figure~\ref{paramHCN}) while \cite{stevenson_etal_2015b} used other sources.
More importantly, together with \cite{stevenson_etal_2015b}, we did not find data which allow the determination of the $k_{ij}$'s for HCN and CH$_3$CN. As emphasized by
\cite{stevenson_etal_2015b} (see their tables 11 and 12) results depend drastically on $k_{ij}$'s; 
this problem could only be solved by future theoretical chemistry simulations or experimental measurements. 
Finally, the ``global picture'' of the evaporite structure does not seem to depend on the chosen model: very soluble
species (like C$_2$H$_2$ and C$_4$H$_{10}$, which are among the most soluble with an RST or PC-SAFT approach) should be predominant at the surface of the central region
of the deposit, whereas less solubles like HCN (all models seem to indicate that nitrogen bearing molecules are poorly soluble) should lay along the former shoreline.
For instance, estimates of HCN solubility much smaller than ours must reenforce the tendency we found.}\\

%
%
%

      ``Tholins'', generated in laboratory experiments, have long been proposed as Titan's aerosol analogs 
\citep{khare_etal_1984,sagan_etal_1992}. Numerous experimental works show that they are complex combinations of C-N-H molecules, 
with a molecular weight ranging between $\sim 100$ to $\sim 800$ daltons
\citep{imanaka_etal_2004,mcdonald_etal_1994,sarker_etal_2003,nna_mvondo_etal_2013}. In addition, they have been found
to be very poorly soluble in nonpolar solvents \citep{mckay_1996,coll_etal_2001}. Tholins are macromalocules much larger than
those considered in this study. In the frame of the
PC-SAFT theory, the segment number $m$ increases with the molecular size, whereas the hard-core segment diameter $\sigma$ and
the segment-segment interaction energy parameter $\epsilon/k_{B}$ remain approximately constant. This behavior can be easily 
understood: $m$ represents the number of ``hard spheres'' that are assumed to compose the molecule in question. One can check this
tendency in Table~6 of \cite{tihic_etal_2008}, where the ratio $m/MW$ ($MW$ stands for the molecular weight) stays around $\sim 0.02$
for a collection of polymers. At the same time, $\sigma$ and $\epsilon/k_{B}$ keep values around $\sim 4$ \AA{ } and
$\sim 250$ K respectively.\\
  For a sensivity analysis, we tried increasing progressively the segment number of butane, 
from its standard value $2.63$ to
$7$ ($m= 8$ has lead to a non-convergence of PC-SAFT, due to an unphysical situation), leaving $\sigma$ and $\epsilon/k_{B}$
unchanged. The solubility of butane has been computed in ethane at $90$ K, under $1.5$ bar. We found that the concentration of
butane would fall from $9.14 \times 10^{-2}$ (in mole fraction) to $1.04 \times 10^{-3}$ (in our hypothetical $m= 7$ case). In other words the
solubility is very sensitive to the value of $m$ and decreases by several order of magnitude when $m$ is increased by only a few
units. More sophisticated numerical experiments \citep[\textit{e.g}. in which $m$ is estimated using a group-contribution 
approach for a complex macromolecule, see ][]{tihic_etal_2008} yield a similar conclusion.\\

   As already emphasized in PAP1,
the enthalpy of melting that appears in Eq. (\ref{EQbase}) has a strong influence of the resulting 
solubility.
   This enthalpy can be estimated using a group-contribution method \cite[see for instance][]{joback_1984}; the relevant equation
proposed in Appendix C of \cite{poling_2007} has the following form
\begin{equation}
\Delta H_{m} = -0.88 + \sum_{k} N_{k}(hmk) \times 0.004184
\end{equation}
where $N_{k}$ is the number of groups (-CH$_3$, -OH, ...) of type $k$, and the $hmk$'s represent the corresponding contributions
to the enthalpy; they are provided by dedicated tables. For the majority of these groups --- 
first of all CH$_3$, CH$_2$ and CH ---
the $hmk$'s are positive. Thus the general tendency is an increase of $\Delta H_{m}$ with the size of molecules 
leading to 
lower solubility. In PAP1, the authors discussed the case of the most simple molecule of the hydrazine family; 
\cite[identified by][to be one of the possible components of tholins]{quirico_etal_2008} CH$_3$CH$_3$N-CH$_2$; and also cyanoacetylene
HC$_3$N. Unfortunately, for both molecules, we were not be able to find in the literature values of even estimations for their PC-SAFT
parameters. Then, we cannot, for this moment, improve the solubility estimations computed in PAP1 for this
particular species.\\ 
  However, Titan's atmospheric aerosols that fall on its surface \citep[][]{barth_toon_2006,larson_etal_2014} --and that could be
similar to tholins-- have probably a very low solubility. Therefore, a layer of these aerosols may compose the lakebed, and most
likely below the layers of butane and acetylene and those of HCN and CH$_3$CN. The ``tholins'' might be buried at the bottom of the
evaporitic deposition or compose the external part of the bathtub.\\
 Present knowledge of the Titan's surface chemical composition suffers from a lack of data. 
\cite{brown_etal_2008} published clues in favor of the presence of the ethane in Ontario Lacus. In their work, \cite{clark_etal_2010} identified
benzene but they could not disentangle spectral signature of HC$_3$N and CO$_2$ while C$_2$H$_2$ has not been 
detected and CH$_3$CN could
explain some spectral features. Beyond this, we have to keep in mind that \textit{in situ} exploration could bring some surprises:
for instance, in their experimental work, \cite{vu_etal_2014} and \cite{cable_etal_2014} have explored the formation of benzene-ethane co-crystals.
Moreover, even if the low temperature of the surface disfavors the kinetics of chemical reactions, cosmics-ray particles could
penetrate down to the surface \cite[][]{sagan_thompson_1984,zhou_etal_2010} and their energy deposition \cite[see][]{molina-cuberos_etal_1999}
could speed-up some simple organics processes that could lead to the emergence
of unexpected species over geological timescales. In short, molecules taken into account in this work as evaporites
are supported by photochemical models, but we cannot exclude that future investigations or possible \textit{in situ} exploration will 
reveal chemical surprises.\\

%
%

   As emphasized in PAP1, \REVsec{possible} turbidity is a major issue in our context. The presence of impurities may play a role
in the nucleation of precipitating organics \REVsec{providing favorable nucleation sites}, and \REVsec{they may} also contribute to the deposits left on the 
ground after evaporation of the liquid.
The authors of PAP1 have recalled that the laws of thermodynamics favour the heterogeneous nucleation \REVsec{since the cost in energy
is lower in the case of a heterogeneous process compared to a homogeneous one}, but
\cite{malaska_hodyss_2013} observed, in their experiments, volume precipitation of benzene (M. Malaska, private communication)
suggesting homogeneous precipitation. This unexpected phenomenon could be explained either by the presence of impurities in the
liquid or by a very smooth internal surface of the experimental cell.
  Among our few observational constraints, \cite{brown_etal_2008}, using observation through the $5$-$\mu$m window, have noticed that Ontario
Lacus appears to be filled with a liquid free of particles larger than a few micrometers. In addition, in their study of Ligeia
Mare \cite{mastrogiuseppe_etal_2014} have also pointed out that the low attenuation of the RADAR signal is compatible with a ternary mixture
of nitrogen, methane and ethane, excluding, after discussion, the possibility of significant turbidity caused by suspended scatterers. Despite
this, the question of turbidity remains an open issue that could be answered by future space missions.\\

%
%
%
   In their study of the geographical distribution of evaporite candidates, \cite{macKenzie_etal_2014} noticed a clear lack of
$5$-$\mu$m-bright material in the south pole district. \REVsec{This observation cannot be explained by water snow falling from the atmosphere
since Titan's atmosphere is particularly poor in water \citep[][]{coustenis_etal_1998,dekok_etal_2007,moreno_etal_2012,cottini_etal_2012}}.
Therefore, \cite{macKenzie_etal_2014} have proposed three possible explanations to this observation:
(1) the evaporites layers could have been buried; being covered by a cap of aerosols, (2) the liquid could have percolated to a
subsurface reservoir through a porous regolith, or (3) that there just haven't ever been deep, long-lived liquids at the south pole.
However, the explanation offered by aerosols settling appears much more unlikely than
percolation, since that implies a probable disappearance of difference between south zones which seems to be water-ice rich, and those
consistent with water-ice in VIMS data.\\
  Two alternative origins of this lack of southward evaporites can be proposed. First, if organic solutes are
mainly produced in the atmosphere, one can imagine the existence of a low production rate above the south polar regions, phenomenon
caused by the photochemistry itself and/or by properties of the atmospheric circulation that could disfavor the south pole. 
\REVsec{This hypothesis is supported by the asymmetrical distribution of lakes, which could be explained by the insolation asymmetry caused by
Saturn's system orbital properties \cite[][]{aharonson_etal_2009,lora_mitchell_2015}.}
Second, due to the high concentrations at saturation of potential solutes into ethane (see Table~\ref{saturationSOLUB}), this solvent,
\REVsec{by running off over the surface},
could \REVsec{dissolve and trap} almost
the entire amount of solid organics that have been fallen from the atmosphere over the south polar region. 
The fact that Ontario
Lacus is recognized to be enriched in ethane \cite[][]{brown_etal_2008,luspay_kuti_etal_2015} strengthens this interpretation.
Moreover, the RADAR very low loss tangent observed in Ligeia Mare by \cite{mastrogiuseppe_etal_2014}, contrasts with the much stronger absorption
estimated at Ontario Lacus \citep{hayes_etal_2010}; this fact could be interpreted as the consequence of the solvation of more absorbing
compounds like long chain hydrocarbons, aromatics and nitriles \citep{mastrogiuseppe_etal_2014}.
Order of magnitude evaluations can be also invoked. For that purpose, we employed the least-square fit, established by
\cite{lorenz_etal_2008}, that provides the average depth of the Earth's $20$ largest lakes, as a function of their size.
(\textit{i.e.} the square root of their surface area). Applied to Ontario Lacus, this law suggests an average depth of
$\sim 200$ m, 
for an adopted surface area of $15,600$ km$^{2}$; finally leading to an approximate total volume of
$\sim 4 \times 10^{12}$ m$^{3}$. If we assume a content made of pure ethane, this volume corresponds to $\sim 10^{17}$ mol, mole
fractions at saturation gathered in Table~\ref{saturationSOLUB} allow estimations of maximum quantities of solutes that can be
contained in Ontario. On another side, we have estimated the total quantities of organic solids settled to the regions further  
south than Ontario Lacus, which represent a total area of about $2 \times 10^{12}$ m$^{2}$; this, during a period equal to a
Titan's year. To do so, we simply have multiplied the rates coming from \cite{lavvas_etal_2008b}, by both the considered area and the
chosen period of time. The results show that the maximum dissolved quantities, allowed by our model, exceed by a factor of
$\sim 10^{3}-10^{4}$ the amounts of potential solutes that are assumed to fall from the atmosphere according to photochemistry
models.
Thanks to Cassini RADAR data, \cite{ventura_etal_2012} derived a more realistic average depth for Ontario Lacus, around ten times lower than our
crude estimation. Obviously, even a much shallower lake does not alter our conclusion.
  Therefore, our scenario, postulating that the south polar regions are currently dominated by liquid ethane that could host the major part
of soluble species, appears plausible.
%
%
%
\section{\label{concl}Conclusion}
  We have developed a new model of dissolution based on the up-to-date theory called PC-SAFT. This model takes into account recent
laboratory measurements. As a by-product, we have determined the PC-SAFT parameters for HCN. The absence of available interaction
parameters $k_{ij}$ for some species, among them HCN, encourages further experimental work on the solubility determinations in
cryogenic solvents.\\
   With our model, we have also computed the possible vertical structure of evaporite deposits. These 1-D simulations confirm
the result already published in PAP1: butane and acetylene are good candidates for species that could compose the surface of evaporite.
In addition, we found that a couple of compounds could form a thick external layer; and due to the combination of the existence of two
crystallographic phases and of the rather thick layer, this external C$_4$H$_{10}$-C$_2$H$_2$ enriched layer could explain the RADAR
brightness of evaporites, \REVsec{if the scale of the produced heterogeneities is similar or larger than the RADAR wavelength}.
We have also shown that the \REVsec{seasonal} cycle may offer a mechanism which
leads to a growth of evaporite thickness only limited by atmospheric production of organics. Thanks to our solubility calculation, we also 
suggest that ethane-enriched south pole lake Ontario Lacus could  have trapped a large quantity of solutes, and this would explain 
--at least partially-- the lack of evaporite in the south polar regions.\\
  Under realistic conditions, with our 2-D model we confirm the possibility of the formation of ``bathtub 
rings'', showing a complex chemical
composition.  However our model suggests the possible existence of ``trimodal
bathtub ring compositions when the entire evaporation is completed. Our predictions are in agreement with
past observations of Ontario Lacus by \cite{barnes_etal_2009a} and \cite{moriconi_etal_2010} and suggest the need of a future Titan's space mission
involving a lander, \REVsec{partly focused on the exploration of the lakes shores, where the chemical diversity} is clearly
high \citep[][]{stevenson_etal_2015}.

%
%
\section*{Abbreviations}
\REVfirst{
\begin{itemize}
  \item PC-SAFT: Perturbed-Chain Statistical Associating Fluid Theory.
  \item VIMS: Visual and Infrared Mapping Spectrometer.
  \item RST: Regular Solution Theory.
  \item EoS: Equation of State.
  \item VLE: Vapor-Liquid Equilibrium.
  \item SLE: Solid-Liquid Equilibrium.
  \item NIST: National Institut of Standards and Technology.
  \item DST: Density Functional Theory.
  \item GCM: Global Circulation Model.
\end{itemize}
}
%
%
\appendix
\section{\label{append}The validity of the equation of liquid-solid equilibrium}

   The solubility calculations, presented throughout this paper, rely on Eq. (\ref{EQbase}), which is --as already mentioned
in PAP1 -- an approximation. The rigorous expression is given by \citep[see for instance, annex of][]{maity_2003}
%
\begin{multline}
\label{rigourousEq}
\mathrm{ln}\,(\Gamma_{i} \, X_{i}^{\rm sat}) = 
   -\underbrace{\frac{\disp\Delta H_{i,m}}{\disp R T_{i,m}} \, \left(\frac{\disp T_{i,m}}{\disp T}-1\right)}_{(1)} \\ 
   -\underbrace{\frac{\disp 1}{\disp R T} \, \int_{P}^{P_{i}^{\rm sat}}\left(V_{i,m}^{S}-V_{i,m}^{L}\right)\, \mathrm{d}P}_{(2)}\\
   -\underbrace{\frac{\disp 1}{\disp R T} \, \int_{T}^{T_{i,m}}\left(C_{p,i}^{S}-C_{p,i}^{L}\right)\, \mathrm{d}T}_{(3)} \\
   +\underbrace{\frac{\disp 1}{\disp T} \, \int_{T}^{T_{i,m}}\left(C_{p,i}^{S}-C_{p,i}^{L}\right)\, 
           \frac{\disp\mathrm{d}T}{\disp T}}_{(4)}     
\end{multline}
%
  and one can legitimately wonder if terms (2), (3) and (4) have a global contribution negligible compared to term (1) or not. It is
very striking to note that experimental data can be nicely reproduced even without these terms (see Fig.~\ref{dotriacontane_heptane} 
and \ref{compaPCSAFTexpe}). It is probably safe then to assume that either terms 2, 3, and 4
have a tiny contribution or that their role is included
in the effect of the interaction parameters $k_{ij}$. Nonetheless, we have tentatively tried to estimate the values of terms
(2)--(4). We found in the literature laboratory measurements for the specific heats $C_{p,i}^{S}$ of some involved solids:
C$_4$H$_{10}$, C$_6$H$_6$, CH$_3$CN and HCN \cite[respectively in ][]{aston_messerly_1940,oliver_etal_1948,putnam_etal_1965,giauque_ruehrwein_1939};
and the specific heats $C_{p,i}^{L}$ of the subcooled liquids can -- at least in principle -- be evaluated by the use of PC-SAFT.
This equation of state only provides their quantities if the Helmholtz energy (or equivalently the specific heat $C_{p,i}^{id}$ of the
corresponding ideal gas) is known. Hence, we have estimated these $C_{p,i}^{id}$ using the group-contribution method developed by
Joback and Reid \cite[][]{joback_1984,joback_reid_1987} and summarized by \cite{poling_2007}. In order to test the validity of this 
approach, we compared the speed of sound
\begin{equation}
c_{\rm sound}= \sqrt{\frac{\disp C_{p}}{\disp C_{v}} \frac{\disp 1}{\disp k_{T}\rho}}
\end{equation}
obtained by this method with experimental results for some cryogenic liquids. The agreement was not good enough to allow a firm
validation of the method. The current development status of our model does not then permit reliable estimation 
for terms (3) and (4).\\
  On one hand, measured molar volumes $V_{i,m}^{S}$ of solids are available (see Table~\ref{table_molvol}); on the other hand the molar
volume $V_{i,m}^{L}$ of subcooled liquids can be computed by PC-SAFT because it is not required to know the Helmholtz energy of the corresponding
ideal gas. The vapor pressures $P^{\rm sat}$ are clearly negligible compared to the ambient pressure of $~\sim 1.5$ bar,
in addition solids and liquids have in general a very low compressibility, that way term (2) in Eq. (\ref{rigourousEq}) is not
significantly different from $\sim |V_{i,m}^{S}-V_{i,m}^{L}| \times P /RT$. Then, for the solid species involved in this work, the
term (1) ranges between $\sim 1$ and $\sim 10$, while $|V_{i,m}^{S}-V_{i,m}^{L}| \times P /RT$ has values of the order of 
$\sim 10^{-4}\textrm{--}10^{-3}$. We conclude that term(2) is negligible, whereas the precise role of the terms (3) and (4) remains questionable,
even if the $k_{ij}$'s could partly mimic their effect.
%


\section*{Acknowledgement}

   We thank Giuseppe Mitri for scientific discussion. TC is funded by the ESA Research Fellowship in Space Science Programme.
The authors acknowledge financial support from the ESAC Faculty (ESAC-358 proposal). Finally, we thank the anonymous Reviewers who 
improved the clarity of the paper with their remarks and comments.

\end{document}